\documentclass[twocolumn]{revtex4}
\usepackage{amsmath}
\usepackage{amssymb}
\usepackage{graphicx,color}
\usepackage[latin1]{inputenc}
\usepackage[T1]{fontenc}
\usepackage{ae,aecompl}

\newcommand{\be}  {\begin{equation}}
\newcommand{\beq} {\begin{equation}}
\newcommand{\bea} {\begin{eqnarray} \nonumber }
\newcommand{\ee}  {\end{equation}}
\newcommand{\eeq} {\end{equation}}
\newcommand{\eea} {\end{eqnarray}}
\newcommand{\beqa}{\begin{eqnarray}}
\newcommand{\eeqa}{\end{eqnarray}}
\newcommand{\bse}{\begin{subequations}}
\newcommand{\ese}{\end{subequations}}

\newcommand{\nn}{\nonumber\\}
\newcommand{\avg}[1]{\langle{#1}\rangle}

\newcommand{\rmd}{\text{d}}

\newcommand{\molecule}[1]{{\mathrm{#1}}}
\newcommand{\MS}{\molecule{S}}
\newcommand{\W} {\molecule{W}}

\newcommand{\X}{\molecule{X}}

\newcommand{\nothing}{\emptyset}

\newcommand{\myav}[1]{\langle{#1}\rangle}

\newcommand{\torate}[1]{\stackrel{#1}{\longrightarrow}}

\newcommand{\power}{S}

\newcommand{\bol}[1]{\mathbf{#1}}
\newcommand{\denom}{\mu^2\lambda^2+(\lambda^2+\mu^2+\nu^2+2\nu(\lambda+\mu))\omega^2+\omega^4}

\newcommand{\tebf}[1]{\mathrm{\mathbf{#1}}}
\newcommand{\give}[1]{\overset{#1}{\longrightarrow}}
\newcommand{\equil}[2]{\underset{#2}{\overset{#1}{\rightleftarrows}}}

\begin{document}

\title{Signal detection, modularity and the correlation between
extrinsic and intrinsic noise in biochemical networks}
\author{
Sorin T\u{a}nase-Nicola$^{1}$, Patrick. B. Warren$^{2}$ and Pieter Rein ten
Wolde$^{1}$}
\affiliation{
$^1$ FOM Institute for Atomic and Molecular Physics, Kruislaan 407,
1098 SJ Amsterdam, The Netherlands\\
 $^2$ Unilever R\&D Port Sunlight, Bebington, Wirral, CH63 3JW, UK.}

\begin{abstract}
Understanding cell function requires an accurate description of how noise is
transmitted through biochemical networks. We present an analytical result
for the power spectrum of the output signal of a biochemical network that
takes into account the correlations between the noise in the input signal
(the extrinsic noise) and the noise in the reactions that constitute the
network (the intrinsic noise). These correlations arise from the fact that
the reactions by which biochemical signals are detected necessarily affect
the signaling molecules and the detection components of the network
simultaneously. We show that anti-correlation between the extrinsic and
intrinsic noise enhances the robustness of zero-order ultrasensitive networks
to biochemical noise. We discuss the consequences of the correlation
between extrinsic and intrinsic noise for a modular description of noise
transmission through large biochemical networks in the context of the
mitogen-activated protein kinase cascade.
\end{abstract}
\date{\today}
\maketitle

Living cells process information in a stochastic manner~
\cite{McAdams97,Elowitz00,Ozbudak02,Elowitz02,Ozbudak04,Pedraza05,Rosenfeld05,
Blake03,Raser04,Acar05,Becksei05,ColmanLerner05,Korobkova04}. It it
generally believed that biochemical noise can be detrimental to cell
function, because it can hamper the reliable integration and
transmission of signals~\cite{Rao02}. It is increasingly becoming
recognized, however, that stochasticity can also play a beneficial
role~\cite{Kaern05}. Biochemical noise can enhance the functioning of
biochemical networks by, \emph{e.g.}, increasing the sensitivity
\cite{Paulsson00_1} or by driving oscillations~\cite{Vilar02}, while
stochastic switching between phenotypes can increase the proliferation
of an organism in a randomly fluctuating
environment~\cite{Thattai04,Kussell05}. It is thus important to
understand how living cells process noisy signals. The complete
biochemical network of a living cell consists of a huge number of
biochemical reactions. This precludes a detailed mesoscopic
description of the propagation of noise through the full
network. However, it is believed that  biochemical networks are modular in design, which
means that they can be decomposed into smaller, functionally
independent units~\cite{Hartwell99,Kashtan05}. This is potentially
useful, because it would make it possible to coarse-grain the full
network of individual reactions to a smaller network consisting of
modules, where each module is described as a `black box' with input
and output signals; the reactions that constitute each module would
then be integrated into the input-output
relations~\cite{Angeli04}. This would allow a simplified description
of the behavior of the full network~\cite{Hartwell99,Angeli04}. It is
not clear, however, whether such a description is possible for the
transmission of noise. Here we address the question: under which
conditions can modularity be exploited to develop a coarse-grained
description for the transmission of noise?

Recently, several groups have derived analytical expressions for the
noise in the output signal of a network as a function of the noise in
the input signal, known as the ``extrinsic noise'', and the noise in the
biochemical reactions that constitute the network, the ``intrinsic
noise''~\cite{Detwiler00,Thattai02,Paulsson04,Shibata05}. These results
suggest that the input-output relations for the noise of the
individual modules of a network can be combined in a simple way to quantify the
transmission of noise through the full
network~\cite{Detwiler00,Thattai02,Shibata05}. However, these studies
assume that the extrinsic and intrinsic noise are independent sources
of noise~\cite{Detwiler00,Thattai02,Paulsson04,Shibata05}. Here, we
show that the biochemical reactions that allow a module to
\emph{detect} the incoming signals introduce correlations between the
fluctuations in the input signals and the intrinsic noise of the
processing module; the extrinsic and intrinsic noise are not
therefore independent of one another. This means that the fluctuations in the
output signal of one module (which depend upon the intrinsic noise of
that module) can affect the intrinsic noise of another module. As a
consequence, the modules do not perform independently of one another
and this, in general, impedes a quantitative modular description of
the propagation of noise through large scale biochemical networks. Our
analysis also reveals the conditions under which detection reactions
do not introduce correlations between the extrinsic and intrinsic
noise; under these conditions, a modular description of noise
transmission can be developed.

\begin{figure*}[t]
\centering \includegraphics[width=14cm]{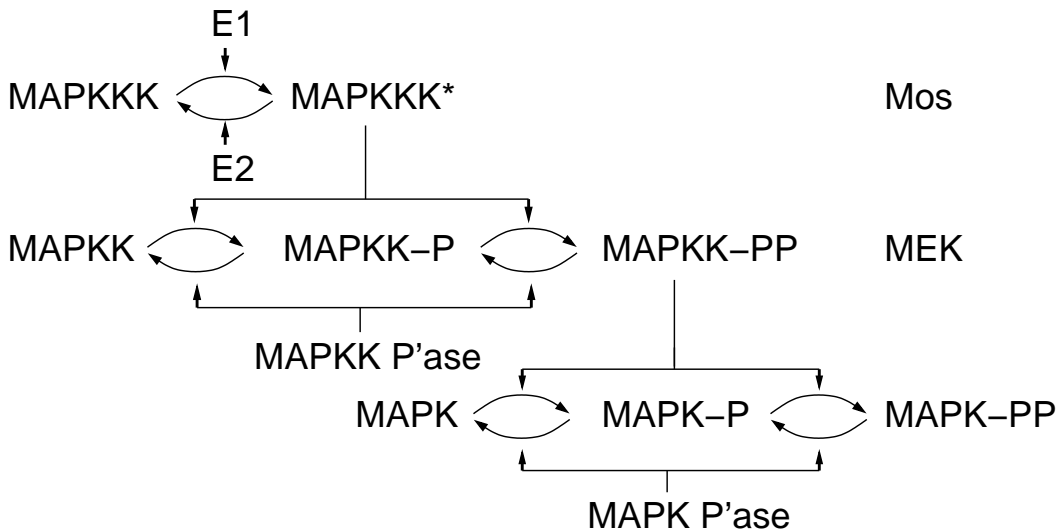}
\caption{The Mos/MEK/p42 MAPK cascade. This network consists of three
  tiers, which, from a topological point of view, could be regarded as
  modules. Note also that the full network consists of push-pull
  networks, in which the activity of an enzyme is covalently modified
  by the action of two opposing enzymes.}
 \label{fig:MAPK}
\end{figure*}

We study the consequences of the correlations between extrinsic and
intrinsic noise for the mechanism of zero-order
ultrasensitivity~\cite{Goldbeter81}. Zero-order ultrasensitivity is
one of the principal mechanisms that allow cells to strongly amplify
input signals~\cite{Goldbeter81,Ferrell96}. It operates in push-pull
networks, which are ubiquitous in prokaryotes and eukaryotes. In a
push-pull network, two enzymes covalently modify a
component in an antagonistic manner, \emph{e.g.}, a kinase that
phosphorylates a component and a phosphatase that dephosphorylates the
same component (see Fig.~\ref{fig:MAPK}). If both enzymes operate near
saturation, then the modification and demodification reactions will be
zero-order, which means that their reaction rates become insensitive
to the substrate concentrations. Under these conditions, a small
change in the concentration of one of the two enzymes, will lead to a
large change in the concentration of the modified protein. Zero-order
ultrasensitivity thus allows push-pull networks to turn a graded input
signal (the concentration of one the enzymes) into a nearly binary
output signal (the covalently modified protein)
\cite{Goldbeter81,Ferrell96}.
However, if the enzymes operate near saturation, then these networks
are also known to exhibit large intrinsic
fluctuations~\cite{Berg00}. This will not only weaken the sharpness of
the macroscopic response curve~\cite{Berg00}, but will also strongly limit
the detectability of the input signal~\cite{Detwiler00}. Moreover,
large extrinsic fluctuations in, \emph{e.g.}, the activity of the kinase, can
induce bistability in the network~\cite{Samoilov05}. In order to
understand the performance of zero-order ultrasensitive networks, it
is therefore important to understand not only their macroscopic
input-output curves, but also their noise characteristics.

The noise properties of zero-order ultrasensitive networks have been
studied
extensively~\cite{Detwiler00,Berg00,Shibata05,Samoilov05}. However,
all these studies have ignored the correlations between the extrinsic
and intrinsic noise: the modification and demodification reactions
were assumed to be given by Michaelis-Mention kinetics without
explicitly taking into account the binding and unbinding of the
enzymes to their
substrates~\cite{Detwiler00,Berg00,Shibata05,Samoilov05}. Here, we
show that the (un)binding of the enzymes to their substrates induces
\emph{anti}-correlations between the extrinsic noise -- the
fluctuations in the activating enzyme -- and the intrinsic noise --
the fluctuations in the modification and demodification
reactions. These anti-correlated fluctuations \emph{reduce} the noise
in the output signal (the covalently modified protein). While it has
long been known that temporal fluctuations in the network components
can adversely affect the performance of zero-order ultrasensitive
networks~\cite{Rao02,Berg00}, our calculations reveal that
anti-correlated fluctuations between different sources of noise can,
in fact, enhance their performance by increasing the signal-to-noise
ratio.

Finally, we illustrate the consequences of the correlations between
extrinsic and intrinsic noise for a modular description of noise
propagation using the MAPK cascade. This network is an important
intracellular signaling pathway in eukaryotes, where it plays a
central role in cell proliferation, cell differentiation, and cell
cycle control~\cite{Ferrell96}. The MAPK cascade consists of three
push-pull modules connected in series (see Fig.~\ref{fig:MAPK}). We
find that because of the correlations, the modules no longer perform
independently of one another. As a result, an iterative application of
noise-input-output relations of the individual modules drastically
overestimates the propagation of noise. Only by analyzing the modules
together at the level of the individual biochemical reactions do we
arrive at the correct result. 

The noise properties of genetic networks in both
prokaryotes~\cite{Elowitz00,Ozbudak02,Elowitz02,Pedraza05,Rosenfeld05}
and eukaryotes~\cite{Blake03,Raser04,Acar05,Becksei05,ColmanLerner05}
have been measured. While noise in signaling pathways has been studied
to some extent in prokaryotes~\cite{Korobkova04}, noise in
 eukaryotic signaling networks has not yet been
experimentally investigated. Our results provide a general,
quantitative framework for describing the transmission of noise in
gene regulatory networks and signal transduction pathways.

\section{Noise addition rule: uncorrelated extrinsic
and intrinsic noise}
We consider a module with one input and
one output signal. We imagine that the module is a `black box': we
will not therefore consider the biochemical reactions that constitute
the module explicitly. Clearly, in general, the incoming signal could
be transformed into the outgoing signal in a complicated manner,
depending upon the reactions that form the module.  We imagine,
however, that the system is in steady state and we assume that the
fluctuations of the incoming and outgoing signals around their
steady-states values are small; this allows us to linearize the
coupling between them and to use the linear-noise
approximation~\cite{VanKampen92}.  Moreover, we will here assume
that the {\em noise} in the input and output signals is {\em uncorrelated}
and that the output signal relaxes exponentially with decay rate
$\mu$. This yields the following chemical Langevin equation for the
output signal
\begin{equation}
\dot{x}=\nu s(t)- \mu x +\eta (t).  \label{eq:addle}
\end{equation}
Here, $s=S-\langle {S}\rangle $ is the deviation of the number of signaling
molecules, $S$, from its mean, $\langle {S}\rangle $, and $x=X-\langle {X}
\rangle $ is the corresponding quantity for the output signal; $\nu$
corresponds to the differential gain and the dot
denotes a time derivative. The last term, $\eta (t)$, describes the
fluctuations in the reactions that constitute the processing unit; we
repeat that we make the crucial assumption that this is uncorrelated from the
input signal $s(t)$. We model $\eta(t)$ as Gaussian white noise: $\langle {\eta (t)}
\rangle =0$ and $\langle {\eta (t)\eta (t^{\prime })}\rangle =\sigma _{\eta
}^{2}\delta (t-t^{\prime })$. By taking the Fourier transform we obtain the
power spectrum $S_{{\mathrm{X}}}$ for the the outgoing signal: 
\begin{equation}
S_{{\mathrm{X}}}(\omega )=\langle {|\tilde{x}(\omega )|^{2}}\rangle =\frac{
2\,\sigma _{\mathrm{in}}^{2}\mu}{\nu^2 + \omega ^{2}}+\frac{\nu ^{2}}{
\mu^{2} + \omega ^{2}}S_{{\mathrm{S}}}(\omega ),  \label{eq:Sadd}
\end{equation}
where $\sigma _{\mathrm{in}}^{2}=\sigma _{\eta }^{2}/(2 \mu)$ is the
intrinsic noise and $S_{{\mathrm{S}}}(\omega )=\langle {|\tilde{s}(\omega
)|^{2}}\rangle $ is the power spectrum of the incoming signal. A similar
expression has been obtained recently~\cite{Detwiler00,Simpson04,Shibata05}.
We refer to Eq.~\ref{eq:Sadd} as the spectral addition rule.

Eq.~\ref{eq:Sadd} has the attractive interpretation that the
computational module acts as a low-pass filter for the input noise
(second term) and generates its own noise in the process (first term);
the filter function $\nu^2/(\mu^2 + \omega^2)$ filters the
high-frequency components in the input signal. Moreover, in this
model, the effects are additive: the intrinsic noise is simply that
which would arise if the input signal did not fluctuate; conversely,
the extrinsic noise is not affected by the intrinsic noise. The
spectral addition rule is a consequence of the linear response of $x$
to the sum $\nu s(t)+\eta(t)$ (Eq.~\ref{eq:addle}), and the assumption that
$s(t)$ and $\eta(t)$ are \emph{uncorrelated}.

If the autocorrelation function for the noise in the incoming signal has an
amplitude $\sigma_{\mathrm{s}}^2$ and decays exponentially with a relaxation
rate $\lambda$, then its power spectrum is given by 
\begin{equation}
S_{{\mathrm{S}}} (\omega) = \frac{2 \, \sigma_{\mathrm{s}}^2\lambda} {
\lambda^{2} + \omega^2 } .  \label{eqsf1}
\end{equation}
In this case, the total noise in the outgoing signal, $\sigma_{\mathrm{tot}
}^2 = 1/(2\pi) \int_{-\infty}^{\infty} d \omega \,S_{{\mathrm{X}}}(\omega)$,
is given by 
\begin{equation}  
\label{eq:noiseadd}
\sigma_{\mathrm{tot}}^2 = \sigma_{\mathrm{in}}^2 + g^2 \frac{\langle{X}
\rangle^2}{\langle{S}\rangle^2} \frac{\mu}{\lambda + \mu} \, \sigma_{\mathrm{
s}}^2 \equiv \sigma_{\mathrm{in}}^2 + \sigma_{\mathrm{ex }}^2.
\end{equation}
Here, $\sigma^2_{\mathrm{ex}}$ is the extrinsic noise and $g\equiv
\partial \ln \langle{X}\rangle  /  \partial \ln \langle{S}\rangle
= (\nu/\mu) \,\, \langle{S}\rangle/ \langle{X}\rangle$ is the
logarithmic gain. This relation has been derived recently by Paulsson~\cite{Paulsson04} and Shibata and Fujimoto~\cite{Shibata05}. We refer to
Eq.~\ref{eq:noiseadd} as the noise addition rule. It is
only valid if the incoming signal has a single exponential relaxation
time and if the spectral addition rule, Eq.~\ref{eq:Sadd}, holds,
which means that the incoming signal and the intrinsic noise must be
uncorrelated.

Eqs.~\ref{eq:Sadd} and~\ref{eq:noiseadd} are potentially powerful
relations, because they allow a modular description of noise
propagation. If, for instance, the network consists of a number of
modules connected in series, then, once the intrinsic noise for each
of the individual modules is known, the propagation of noise through
the network can be obtained for\emph{\ arbitrarily varying} input
signals by the recursive application of the spectral addition rule,
Eq.~\ref{eq:Sadd}, to the successive
modules~\cite{Detwiler00,Thattai02}; the noise strength of the output
signal of the network could then be obtained by integrating over its power
spectrum. However, this approach requires that
the spectral addition rule holds for each of the individual
modules. Below, we will show that the detection reactions can
introduce correlations between the extrinsic and intrinsic
noise. These correlations lead to a break down of the spectral
addition rule, impeding a quantitative modular description of the
transmission of noise through large networks.

\section{Correlated extrinsic and intrinsic noise}
\label{sec:CorExIn}
 To elucidate the origin of the correlations between the
extrinsic and intrinsic noise, it will be instructive to consider a
module that consists of one component only. This component,
$\mathrm{X}$, both detects the input signal $\mathrm{S}$ and provides
the output signal. In the next section we will generalize our results
and discuss more complex modules. To capture the
correlations between the extrinsic and intrinsic noise, we explicitly
describe the detection of the signal by studying the \emph{coupled}
Langevin equations for the two interacting species --  the signaling
molecules $\mathrm{S}$ and the detection molecules $\mathrm{X}$; only
by analyzing the input signal and the processing module together, do
we arrive at the correct results.

\begin{table*}[t]
\begin{tabular}{lll}
 & {}\qquad Scheme & {}\qquad Examples \\[3pt]
\hline\\[-6pt]
(I)& $\MS + \W  \overset{\nu=k_f \W}{\underset{\mu}{\rightleftarrows}}\X$&
\begin{minipage}{9.0cm}\flushleft
binding of ligands to receptors\\
binding of enzymes to substrates\\
binding of transcription factors 
to operator sites
\end{minipage}
\\[15pt]
\hline\\[-6pt]
(II)&
$\MS\torate{\nu}\X\torate{\mu}{\nothing}$ &
\begin{minipage}{9.0cm}\flushleft
post-translational modification\\
activation of trans-membrane 
receptors followed by endocytosis
\end{minipage}
\\[9pt]
\hline\\[-6pt]
(III) \, &
$\MS\torate{\nu}\MS+\X$,\,
$\X\torate{\mu}{\nothing}$ \, &
\begin{minipage}{9.0cm}\flushleft
coarse-grained models 
of enzymatic reactions\\
regulation of gene expression
\end{minipage}
\\[9pt]
\hline
\end{tabular}

\caption[?]{ Three elementary detection motifs. Here, ${\mathrm{X}}$ is the output
signal and ${\mathrm{S}}$ is the input signal; in motif I, ${\mathrm{W}}$ is
the inactive (unbound) state of the detection component and ${\mathrm{X}}$
the active (bound) state. All these schemes have ${\emptyset}\overset{k}{
\longrightarrow}n\,{\mathrm{S}}$ and ${\mathrm{S}}\overset{\lambda}{
\longrightarrow}{\emptyset}$ as a precursor to generate an input signal
which fluctuates about a steady state. For all the schemes, the mean output
signal obeys $d\langle{X}\rangle/dt=\nu\langle{S}\rangle-\mu\langle{X}
\rangle $. A class of examples of detection motif II is the endocytosis of
receptors such as the EGF receptors, which, upon ligand binding, are
endocytosed together with the ligand~\cite{Felder92}.}
 \label{tab:detmotif}

\end{table*}

The most general form of the two coupled chemical Langevin equations is 
\begin{subequations}
\label{eq:gle}
\begin{align}
\dot{s}& =-\lambda s+\kappa x+\xi (t), \\
\dot{x}& =\nu s-\mu x+\eta (t).
\end{align}
Here, $\kappa $ describes how the number of signaling molecules is directly
affected by the number of (active) detection molecules and $\xi (t)$ models
the noise in $s$; the noise terms, $\xi (t)$ and $\eta (t)$, are modeled as 
\emph{correlated} Gaussian white noise: 
$\langle {\xi (t)}\rangle =\langle {\eta(t)}\rangle =0$, 
$\langle {\xi (t)\xi (t^{\prime })}\rangle=\langle {\xi ^{2}}\rangle \delta (t-t^{\prime })$, 
$\langle {\eta(t)\eta (t^{\prime })}\rangle =\langle {\eta ^{2}}\rangle \delta
(t-t^{\prime })$ and $\langle {\xi (t)\eta (t^{\prime })}\rangle =\langle {
\xi \eta } \rangle \delta (t-t^{\prime })$. The pair of linear
differential equations in Eq.~\ref{eq:gle} can be solved using Fourier transformation, which
leads to the following power spectra for the signaling molecules, $S_{{
\mathrm{S}}}(\omega )$, and detection molecules, $S_{{\mathrm{X}}}(\omega )$
, 
\end{subequations}
\begin{subequations}
\label{eq:Sgle}
\begin{align}
S_{{\mathrm{S}}}(\omega )& ={\frac{{\kappa ^{2}}\,{\langle {\eta ^{2}}
\rangle }+2\,\kappa \,\mu \,\langle {\xi \eta }\rangle +\left( {\mu ^{2}}+{{
\omega }^{2}}\right) \,{\langle {\xi ^{2}}\rangle }}{{{\left( \kappa \,\nu
-\lambda \,\mu \right) }^{2}}+\left( {\lambda ^{2}}+2\,\kappa \,\nu +{\mu
^{2}}\right) \,{{\omega }^{2}}+{{\omega }^{4}}}}, \\
S_{{\mathrm{X}}}(\omega )& ={\frac{\langle {\eta ^{2}}\rangle \,\left( {
\lambda ^{2}}+{{\omega }^{2}}\right) +2\,\lambda \,\nu \,\langle {\xi \eta }
\rangle +{\nu ^{2}}\,\langle {\xi ^{2}}\rangle }{{{\left( \kappa \,\nu
-\lambda \,\mu \right) }^{2}}+\left( {\lambda ^{2}}+2\,\kappa \,\nu +{\mu
^{2}}\right) \,{{\omega }^{2}}+{{\omega }^{4}}}}.  \label{eq:SgleX}
\end{align}
\end{subequations}
In contrast to Eq.~\ref {eq:Sadd}, Eq.~\ref{eq:Sgle} takes
into account the correlation between the noise in the external signal
and the intrinsic noise of the processing unit.  Importantly, if both
$\kappa $ and $\langle {\xi \eta }\rangle $ are zero, then
Eq.~\ref{eq:Sgle} reduces to Eq.~\ref{eq:Sadd}.

The correlations between extrinsic and intrinsic noise can have two
distinct origins. Both are a consequence of the molecular character of
the components; the correlations are thus unique to biochemical
networks and absent in electronic circuits. We will illustrate the
sources of correlations using
three elementary detection motifs, which are described in 
Table~\ref{tab:detmotif}; Table~\ref{tab:react} shows their
power spectra and noise strengths. The motifs obey the same
macroscopic chemical rate equation and have the same intrinsic
noise. However, the noise in the input signals is transmitted
differently. The differences between the motifs are due to the
different sources of correlations that are present.

The first source of correlation between the extrinsic and intrinsic
noise arises from the fact that the processing unit can act back on
the input signal by directly affecting the number of signaling
molecules; this corresponds to a non-zero value of $\kappa$. This
source of correlation is present in detection motif I, where it arises
from the unbinding of signaling molecules from the detection
molecules. For scheme I, $\kappa = \mu$; the effect of this source of
correlation on the noise in $\mathrm{S}$ and $\mathrm{X}$ depends upon
the values of all the rate constants (see Eq.~\ref{eq:Sgle}) and can
either be negative or positive. The second possible source of
correlation between the extrinsic and intrinsic noise is the
\emph{correlated fluctuations} in the number of signaling molecules
and detection molecules, which arise from detection reactions that
\emph{simultaneously} change the number of both species; these
correlated fluctuations are quantified via the cross-correlation
function $\langle {\xi (t)\,\eta (t^{\prime })}\rangle$. This source
of correlation is present in both schemes I and II, since each time a
detection reaction fires, a signaling molecule is consumed and
simultaneously a molecule of the processing module is produced
(or activated). Additionally, for scheme I, the unbinding reactions
also lead to cross-correlations in $\xi(t)$ and $\eta (t)$. For scheme I,
$\langle{\xi\eta}\rangle = - (\nu \langle{S}\rangle + \mu \langle{X
}\rangle)$, while for scheme II, $\langle{\xi\eta}\rangle = - \nu
\langle{S} \rangle$~\cite{Gillespie00}. We emphasize that the sign of
$\langle{\xi\eta} \rangle$ is negative; the extrinsic and intrinsic
noise are thus anti-correlated. Eq.~\ref{eq:Sgle} shows that these
anti-correlated fluctuations can lower the noise in $\mathrm{S}$ and
$\mathrm{X}$.

\begin{table*}[t]

\begin{tabular}{cll}
Scheme & {}\qquad Noise strength &
 {}\qquad Noise power spectrum \\[3pt]
\hline\\[-6pt]
(I) &
$\begin{array}{l}
\displaystyle{\sigma_{\mathrm{s}}^2=\myav{S}\,
\frac{(n+1)(\lambda+\mu) + 2\nu}{2(\lambda+\mu+\nu)}} \\[12pt]
\displaystyle{\sigma_{\mathrm{x}}^2=\sigma_{\mathrm{in}}^2+
g^2 \frac{\myav{X}^2}{\myav{S}^2}
\frac{(n-1)\mu}{(n+1)(\lambda+\mu)+2\nu}\,\sigma_{\mathrm{s}}^2}
\end{array}$ &
$\begin{array}{l}
\displaystyle{\power_{\mathrm{s}}(\omega)=\myav{S}\,
\frac{(n+1)\mu^2\lambda+(2\nu+(n+1)\lambda)\omega^2}{\denom}} \\[12pt]
\displaystyle{\power_{\mathrm{x}}(\omega)=
\myav{X}\,\frac{\mu[\lambda(2\lambda+(n+1)\nu)+2\omega^2]}{\denom}}
\end{array}$ \\[24pt]
\hline\\[-6pt]
(II) &
$\begin{array}{l}
\displaystyle{\sigma_{\mathrm{s}}^2=\myav{S}\,\frac{n+1}{2}} \\[12pt]
\displaystyle{\sigma_{\mathrm{x}}^2=\sigma_{\mathrm{in}}^2+
g^2 \frac{\myav{X}^2}{\myav{S}^2}
\frac{(n-1)\mu}{(n+1)(\lambda+\mu+\nu)}\,\sigma_{\mathrm{s}}^2}
\end{array}$ &
$\begin{array}{l}
\displaystyle{\power_{\mathrm{s}}(\omega)=\myav{S}\,
\frac{(n+1)(\lambda+\nu)}
{(\lambda+\nu)^2+\omega^2}} \\[12pt]
\displaystyle{\power_{\mathrm{x}}(\omega)=
\frac{2\sigma_{\mathrm{in}}^2\mu}{\mu^2+\omega^2} +
\frac{\nu^2}{\mu^2+\omega^2}\,
\frac{n-1}{n+1}\,\power_{\MS}(\omega)}
\end{array}$ \\[24pt]
\hline\\[-6pt]
(III) &
$\begin{array}{l}
\displaystyle{\sigma_{\mathrm{s}}^2=\myav{S}\,\frac{n+1}{2}} \\[12pt]
\displaystyle{\sigma_{\mathrm{x}}^2=\sigma_{\mathrm{in}}^2+
g^2 \frac{\myav{X}^2}{\myav{S}^2}
\frac{\mu}{\lambda+\mu}\,\sigma_{\mathrm{s}}^2}
\end{array}$ &
$\begin{array}{l}
\displaystyle{\power_{\mathrm{s}}(\omega)=\myav{S}\,
\frac{(n+1)\lambda}{\lambda^2+\omega^2}} \\[12pt]
\displaystyle{\power_{\mathrm{x}}(\omega)=
\frac{2\sigma_{\mathrm{in}}^2\mu}{\mu^2+\omega^2} + 
\frac{\nu^2}{\mu^2+\omega^2}\,
\power_{\MS}(\omega)}
\end{array}$ \\[24pt]
\hline
\end{tabular}

\caption[?]{Exact solutions for the noise strengths and the noise
power spectra, for the three detection motifs shown in Table~\ref{tab:detmotif}; in scheme I, the detection molecules $W$ are assumed to be
present in abundance. The noise strengths $\sigma_{\mathrm{s}}^2$ and
$\sigma_{\mathrm{x}}^2$ correspond to the noise strengths of the input and
output signal, respectively, and $S_{{\mathrm{S}}} (\omega)$ and $
S_{{\mathrm{X}}} (\omega)$ denote the respective power spectra. For all
motifs, the intrinsic noise is $\sigma_{\mathrm{in}}^2=
\langle{X}\rangle$ and the logarithmic gain $g=\partial \ln
\langle{X}\rangle \, / \, \partial \ln \langle{S}\rangle = (\nu / \mu)
\,\, \langle{S}\rangle/ \langle{X}\rangle$.  Only scheme III obeys the
spectral and noise addition rules (see Eqs.~\ref{eq:Sadd} and~\ref{eq:noiseadd}). }

\label{tab:react}
\end{table*}

For detection motif III, both $\kappa$ and $\langle{\xi\eta}\rangle$
are zero. This scheme is the only one for which there is no
correlation between the extrinsic and intrinsic noise. In motif III,
the incoming signal catalyzes the activation of detection
molecules. This reaction does not affect the signal in any way. The
extrinsic and intrinsic noise are therefore
uncorrelated and the spectral addition rule, Eq.~\ref{eq:Sadd},
holds. In this example the input signal relaxes mono-exponentially, so
that the noise addition rule, Eq.~\ref{eq:noiseadd}, also
holds.

\section{General processing modules and modular
description of noise transmission}
 The systems discussed in
the previous section consist of one component only. In appendix~\ref{app:GenProc}, we derive the power spectra of the output signals of
modules that consist of an arbitrary number of linear(ized) reactions;
they are a generalization of Eq.~\ref{eq:Sgle}. Here, we summarize the
main results and discuss how correlations between extrinsic and
intrinsic noise affect a modular description of noise transmission
through large networks.

If the incoming signals of a module are uncorrelated with each other
and are processed via detection scheme III, then the power spectrum
of the outgoing signal ${\mathrm{X}}_i$ is given by
\begin{equation}
S_{{\mathrm{X}}_i}(\omega)=S^{\mathrm{in}}_{{\mathrm{X}}_i}(\omega)+
\sum_j \tilde
g_{\mathrm{X}_i}^{\mathrm{S}_j} (\omega)S_{\mathrm{S}_j}(\omega).  \label{eq:genshibata}
\end{equation}
The first term on the right-hand side is the power spectrum of the
intrinsic noise in ${\mathrm{X}}_i$ and $\tilde
g_{\mathrm{X}_i}^{\mathrm{S}_j} (\omega)$ is the frequency dependent
gain corresponding to input signal $\mathrm{S}_j$. The gain $\tilde
g_{\mathrm{X}_i}^{\mathrm{S}_j} (\omega)$ is determined by the
coupling between the network components (see appendix~\ref{app:GenProc}),
which, in general, relax multi-exponentially. A noise addition rule
analogous to Eq.~\ref{eq:noiseadd} is therefore no longer obeyed. A
spectral addition rule, Eq.~\ref{eq:genshibata}, nevertheless still
holds for these modules, because the extrinsic and intrinsic noise are
independent of one another -- the input signals are detected via motif
III. Accordingly, the extrinsic contributions to the power spectrum of
the output signal can be \emph{factorized} into a function that only
depends upon the intrinsic properties of the module, namely $\tilde
g_{\mathrm{X}_i}^{\mathrm{S}_j} (\omega)$, and one that only depends
upon the input signal, $S_{\mathrm{S}_j}(\omega)$. This allows a
simple and modular description of noise transmission.

If a module receives its input via detection scheme I or II, then
correlations will be induced between the noise in the input signals
and the intrinsic noise of the module.  In this case, the spectral
addition rule, Eq.~\ref{eq:genshibata}, and hence the noise addition
rule~\cite{Paulsson04,Shibata05} break down. More importantly, the
correlations mean that the noise in the output signal of one module
(which depends upon the intrinsic noise of that module) affects the
intrinsic noise of another module. As a result, the intrinsic
fluctuations of the different modules become correlated with one
another; the modules therefore no longer perform independently of one
another. This precludes a modular description of noise propagation,
because the modules have to be analyzed together at the mesoscopic
level of the individual biochemical reactions.


\section{Zero-order ultrasensitivity}
We illustrate the consequences of correlated extrinsic and intrinsic noise
using the following push-pull network (see also Fig.~\ref{fig:MAPK}): 
\begin{subequations}
\label{eq:PP}
\begin{align}
{\mathrm{E_a}} + {\mathrm{W}} \overset{a_1}{\underset{d_1}{\rightleftarrows}}
{\ \mathrm{E_a W}} \overset{k_1}{\longrightarrow} {\mathrm{E_a}} + {\mathrm{X
}} \\
{\mathrm{E_d}} + {\mathrm{X}}  \overset{a_2}{\underset{d_2}{\rightleftarrows}}
{\ \mathrm{E_d X}} \overset{k_2}{\longrightarrow} {\mathrm{E_d}} + {\mathrm{W
}}
\end{align}
\end{subequations}
Here, ${\mathrm{E_a}}$ is the activating enzyme that provides the
incoming signal and ${\mathrm{E_d}}$ is the deactivating enzyme; the
substrate ${\ \mathrm{W}}$ is the unmodified component that serves as
the detection component and ${\mathrm{X}}$ is the modified component
that provides the outgoing signal. 

We have computed the noise in the output signal ${\rm X}$ for the
push-pull network of Eq.~\ref{eq:PP}. The input signal ${\rm E_a}$ is
modeled as a birth-death process, corresponding to (de)activation of
${\rm E_a}$. The analysis has been performed using the
linear-noise approximation~\cite{VanKampen92} (see
Appendix~\ref{app:pp}) and its accuracy was verified by
performing kinetic Monte Carlo simulations of the chemical master equation~\cite{Bortz75,Gillespie77}. We
found that the analytical results are accurate to within 10\%. Only at
very high enzyme saturation, $[\rm W_{\rm T}]/[\rm E_{aT}] > 100$, do
the numerical results deviate significantly from the analytical
results of the linear-noise approximation, because, in that regime, the 
fluctuations become 
are very large; this is due to the fact
that, when the enzymes are fully saturated,  the behavior of the
push-pull network resembles that of a system that is close to the
critical point of a thermodynamic phase transition~\cite{Berg00}.

\begin{figure}[t]
\centering \includegraphics[width=8cm]{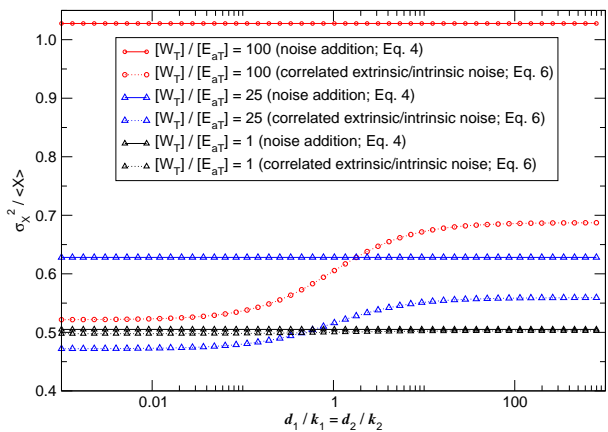}
\caption{The noise of ${\mathrm{X}}$, $\sigma^2_{{\mathrm{X}}} /
\langle{X} \rangle$ as a function of $d_1/k_1 = d_2/k_2$ for a
push-pull network for three different total substrate concentrations
$[\mathrm{W_{\rm T}}] = [\mathrm{W}] + [\mathrm{E_a W}];
[\mathrm{E_{a,T}}] = [\mathrm{E_a}] + [\mathrm{E_a W}]$ (see Eqs.~\ref{eq:PP}). In all cases, the
Michaelis-Menten constants were kept constant at $K_{M,E_a} =
K_{M,E_d} = K_M = 20$ $\mu$M. The horizontal solid lines correspond to the
predictions of the noise addition rule (see Eq.~\ref{eq:noiseadd}),
while the dotted curves correspond to the results of the
full analysis that takes into account the correlations between the
extrinsic and intrinsic noise. Note that as the substrate
concentration increases, the push-pull network moves deeper into the
zero-order regime and the noise in the output signal $\mathrm{X}$
increases. However, the actual increase is significantly lower than
that predicted by the noise addition rule. The smaller increase in the
output noise is due to the anti-correlations between the noise in the
input signal and the intrinsic noise of the push-pull network. The decay
rate of the enzyme $\mathrm{E_a}$ is $\lambda=30 \, k_1$, while the
production rate is chosen such that
$[\mathrm{E_a}] = [\mathrm{E_d}]= 0.2$ $\mu$M.}
\label{fig:noise_pp}
\end{figure}

It is known that the noise in the output signal can increase as the
biochemical reactions by which modules detect incoming signals become
slower~\cite{Bialek05}. We have therefore varied the rate of binding
and unbinding of the enzymes to their
substrates. Fig.~\ref{fig:noise_pp} shows the noise in the output
signal as a function of the enzyme-substrate (un)binding rate, for
different substrate concentrations; the ratios $d_1/k_1$ and $d_2/k_2$
are varied such that the Michaelis-Menten constants $K_{\mathrm{M,a}}
\equiv (d_1+k_1)/a_1 = K_{\mathrm{M,d}} \equiv (d_2 + k_2)/a_2 =
K_{\mathrm{M}}$ remain constant. We find that as the detection
reactions become slower (lower $d_1/k_1$), the noise in the output
signal decreases, rather than increases.  The reason for this, perhaps
counterintuitive, result is that the detection reactions not only
provide a source of biochemical noise, but also act to integrate the
fluctuations in the input signal.

In the limit that both $d_1 \gg k_1$ and $d_2 \gg k_2$, the (un)binding
reactions of the enzymes to the substrates can be integrated out and the
reaction scheme reduces to 
\begin{subequations}
\label{eq:PPadd}
\begin{align}
{\mathrm{E_a}} + {\mathrm{W}} \, {\rightarrow} \, {\mathrm{E_a}} + {\mathrm{X
}} \\
{\mathrm{E_d}} + {\mathrm{X}} \, {\rightarrow} \, {\mathrm{E_d}} + {\mathrm{W
}}
\end{align}
\end{subequations}
In this limit, the signal $\mathrm{E_a}$ is detected via scheme III
(see Table~\ref{tab:detmotif}); the modification and demodification
rates are then given by Michaelis-Menten
formulae~\cite{Shibata05}. However, even in this limit, the spectral
addition rule does not necessarily apply (see
Fig.~\ref{fig:noise_pp}), as has been
suggested~\cite{Shibata05}. Because the total amount of $\mathrm{W}$
and $\mathrm{X}$ together is conserved, fluctuations in $\mathrm{X}$
are anti-correlated with those in $\mathrm{W}$ and hence with those in
the fraction of enzyme $\mathrm{E_a}$ that is bound to its substrate
$\mathrm{W}$. Since the bound enzyme is protected
from deactivation, this effect will introduce anti-correlations
between the fluctuations in the input signal $\mathrm{E_{a,T}}$ -- the
extrinsic noise -- and those in the modification and demodification
reactions -- the intrinsic noise (see appendix~\ref{app:pp}). This
means that, even when $d_1 \gg k_1$ and $d_2 \gg k_2$, the spectral
addition rule, and hence the noise addition rule, only holds when the
fraction of bound enzyme is negligible (see
Fig.~\ref{fig:noise_pp}). 

It is commonly believed that one of the main biological functions of
push-pull networks is to turn a graded input signal into a nearly
binary output signal, thus allowing for an all-or-none
response~\cite{Goldbeter81,Ferrell96}. The amplification of input
signals in push-pull networks strongly increases with increasing
enzyme saturation~\cite{Goldbeter81}. Fig.~\ref{fig:noise_pp} shows
that as the substrate concentrations are increased (and the enzymes
thus become more saturated with substrate), the noise in the output
signal also markedly increases.  This has been observed
before~\cite{Berg00,Shibata05}: the higher gain not only amplifies the
mean input signal, but also the noise in the input signal, the
extrinsic noise~\cite{Shibata05}; in addition, as the enzymes become
more saturated with substrate, the intrinsic fluctuations of the
network also increase, because the network moves deeper into the
zero-order regime~\cite{Berg00}. Our results show, however, that the actual
increase in the output noise is much lower than that
predicted by the noise addition rule.  This is a result of the
anti-correlations between the fluctuations in the input signal
($\mathrm{E_{a}}$) and the intrinsic noise of the network. These
anti-correlations between the extrinsic and intrinsic noise reduce the
noise in the output signal, but are neglected by the noise addition
rule.  Moreover, they become more important as the network moves
deeper into the zero-order regime -- as the enzymes $\mathrm{E_{a}}$
and $\mathrm{E_{d}}$ become more saturated with their substrates, the
input signal $\mathrm{E_{a}} $ is increasingly being affected by its
interaction with the detection component $\mathrm{W}$ of the
network. While it is known that push-pull networks operating in the
zero-order regime tend to exhibit large intrinsic
fluctuations~\cite{Berg00,Rao02}, our results show that
anti-correlations between the extrinsic and intrinsic noise temper
these fluctuations. 

\begin{table*}[t]
\begin{tabular}{l|c|c|c|}

\mbox{} &
$\sigma^2_{\mathrm{Mos^*}}/{\left[\mathrm{Mos^*}\right]}$  &
$\sigma^2_{\mathrm{MEK-PP}}/{\left[{\mathrm{MEK-PP}}\right]}$ &
$\sigma^2_{\mathrm{MAPK-PP}}/{\left[{\mathrm{MAPK-PP}}\right]}$ \\
\hline\\[-6pt]
Fully coupled ${\mathrm{Mos}}$, ${\mathrm{MEK}}$ and $\mathrm{MAPK}$  & 0.643 & 90.3 & 2.25  \\
\hline\\[-6pt]
Spectral addition rule; all uncoupled & 0.727 & 168. & 3.75\\
\hline\\[-6pt]
Coupled ${\mathrm{Mos}}$ and ${\mathrm{MEK}}$; uncoupled $\mathrm{MAPK}$ &0.643 &91.1  & 2.26 \\
\hline\\[-6pt]
Coupled ${\mathrm{MEK}}$ and ${\mathrm{MAPK}}$; uncoupled $\mathrm{Mos}$ & 0.727&166. & 3.72 \\
\hline
\end{tabular}
\caption{Noise in the Mos/MEK/p42 MAPK cascade (see
Fig.~\ref{fig:MAPK}). ``Full noise'' refers to results of the
analysis, in which the transmission of noise was computed by studying
all the biochemical reactions together. ``Spectral addition''
corresponds to the results of the recursive application of the
spectral addition rule to all the successive layers; the layers are
thus assumed to form independent modules. ``Coupled Mos and MEK;
uncoupled MAPK''
refers to the results of an analysis, in which the first two layers
were considered to form one module, while the third layer was assumed
to form an independent module; the transmission of noise from the
first to the second layer was thus computed by taking into account the
correlations between the fluctuations in the output signal of the
first layer, the concentration of active Mos, and the intrinsic noise
of the second layer, the intrinsic fluctuations of MEK; in contrast,
the transmission of noise from the second to the third module was
calculated using the spectral addition rule, thus ignoring these
correlations. ``Coupled MEK and MAPK; uncoupled Mos'' corresponds to the analysis in
which the first layer was considered to be independent, while the last
two layers were considered to form one module. Note that the
correlations between the noise in active Mos and the intrinsic
fluctuations of MEK strongly affect the transmitted noise. The
concentrations are: [Mos] = 3 nM; [MEK] = 1200 nM; [MAPK] = 300 nM.  See
appendix~\ref{app:MAPK} for more details.}
\label{tab:MAPK}
\end{table*}

\section{Modularity and the MAPK cascade}
 We
show the implications of the correlations between the extrinsic and
intrinsic noise for a modular description of noise transmission in
biochemical networks using a MAPK cascade that is based on the
Mos/MEK/p42 MAPK cascade of \emph{Xenopus} oocytes. From a topological
point of view, this network
consists of three push-pull modules that are connected in series. Their key
features are shown in Fig.~\ref{fig:MAPK}. 

Active Mos can activate MEK through phosphorylation at two residues,
while MEK, in turn, can activate p42 MAPK, also via double
phosphorylation. The output signal of each of the modules, which
provides the input signal for the next module in the cascade, is thus
the active enzyme. We study here the network in which the feedback
from p42 MAPK to Mos has been blocked~\cite{Huang96,Ferrell98}.  The
rate constants and the protein concentrations were, as far as
possible, taken from experiments~\cite{Angeli04,Ferrell96,Huang96} (see
appendix~\ref{app:MAPK} for details). All results have been obtained
using the linear-noise approximations~\cite{VanKampen92}. These analytical
results were found to be within 25\% of numerical results of
kinetic Monte Carlo simulations of the chemical master
equation~\cite{Bortz75,Gillespie77}. Appendix~\ref{app:MAPK} describes in 
detail the linear-noise analysis.

Table~\ref{tab:MAPK} shows the noise in the output signals of the
three modules, as predicted by an iterative application of the spectral
addition rule to the successive modules and as revealed by the
analysis that takes into account the correlations between the
fluctuations in the input signal of a module (which is the output
signal of its upstream module) and the intrinsic noise of that module
(see appendix~\ref{app:MAPK}). The spectral addition rule significantly
overestimates the propagation of noise; the noise in MAPK, the output
signal of the cascade, is about 50\% lower than that predicted by the
spectral addition rule.  This supports our conclusion that
anti-correlations between the extrinsic and intrinsic noise can
enhance the performance of biochemical networks by making them more
robust against biochemical noise.

Table~\ref{tab:MAPK} also illustrates under which conditions
modularity can be exploited for a coarse-grained description of noise
transmission. ``Coupled Mos and MEK; uncoupled MAPK'' refers to the
results of an analysis, in which it is assumed that the first two
layers together form one module, while the third layer constitutes a
second, independent module. This analysis takes into account the
correlations between the fluctuations in the output signal of the
first layer -- the concentration of active Mos -- and the intrinsic
fluctuations of the second layer -- the noise in the modification and
demodification reactions of MEK. However, it ignores the correlations
between the output signal of the second layer and the intrinsic noise
of the third layer. Similarly, ``Coupled MEK and MAPK; uncoupled Mos''
corresponds to the results of a description, in which the first layer
forms one module, whereas the second and third layer form a second,
independent module. We refer to appendix~\ref{app:MAPK} for a
mathematically precise description of the coupling between the
respective layers. 

It is seen that while the ``Coupled Mos and MEK; uncoupled MAPK''
description is fairly accurate, the ``Coupled MEK and MAPK; uncoupled
Mos'' description significantly overestimates the noise in the output
signal of the cascade. This shows that while the correlations between
extrinsic and intrinsic noise are not very important for the
transmission of noise from the second to the third layer, these
correlations do strongly affect the propagation of noise from the
first to the second layer. This difference is due to the differences
in enzyme saturation: active Mos is more saturated with its substrate,
MEK, than active MEK is with its substrate, MAPK. As a consequence,
the signal that connects the first with the second layer (active Mos)
is more affected by the detection reactions than that which connects
the second with the third (active MEK).

\section{Discussion}
This analysis of the
MAPK cascade unambiguously shows that correlations between the noise
in the input signal of a network and the noise in the biochemical
reactions that constitute the network, will affect the noise in the
output signal of that network. It also shows the implications for a
coarse-grained description of noise propagation in biochemical
networks that, from a topological point of view, appear to consist of
functionally independent modules~\cite{Hartwell99,Kashtan05,Angeli04}.

From the perspective of noise transmission, a network can be
decomposed into modules only if the signals that connect them are
detected via chemical reactions that do not introduce significant
correlations between the noise in these signals and the intrinsic
noise of the modules. Each module then has a {\em characteristic}
noise input-output relation (Eq.~\ref{eq:genshibata}) and this admits
a modular description of noise transmission. The Mos-MEK/MAPK
decomposition of the MAPK cascade illustrates this: the transmission
of noise from the Mos-MEK subnetwork to the MAPK subnetwork can be described
at the level of two independent modules connected in series, because
the detection reactions do not affect the signal that connects the
modules.  If, however, the correlations between the extrinsic and
intrinsic noise are significant, then the noise input-output relations of
the different subnetworks become entangled and they no longer perform
independently of one another, as the Mos/MEK-MAPK partitioning
shows. In these cases, the propagation of noise can only be quantified
accurately if the correlated subnetworks are regrouped into
independent modules.

The effect of the correlations between the extrinsic and intrinsic
noise is difficult to predict \textit{a priori}.  Our analytical
expression for the power spectrum of the output signal (see
Eq.~\ref{eq:Sgle}) reveals that the effect on
the noise in the output signal can either be positive or negative.
This means that even if one would like to derive a lower or an upper
bound on the transmitted noise, one has to explicitly take into
account the correlations between the extrinsic and intrinsic
noise. The noise addition rule, which neglects the correlations, makes
uncontrolled assumptions and should thus be used with
care~\cite{Detwiler00,Paulsson04,Shibata05}. 

In general, the importance of the correlations between the extrinsic
and intrinsic noise for the noise transmission is determined by the
extent to which the signal is affected by the detection
reactions. This increases when the concentration of the signaling
molecules decreases with respect to that of the detection components
(see Fig.~\ref{fig:noise_pp}). In zero-order ultrasensitive networks,
the correlations are important, because the enzymes are saturated with
substrate -- the enzyme concentrations are thus low compared to the
substrate concentrations. In gene regulatory networks, correlations
between the fluctuations in the concentrations of the gene regulatory
proteins, the extrinsic noise, and the intrinsic noise of gene
expression, can contribute significantly to the transmission of noise
through the network (to be published), because the concentrations of
the gene regulatory proteins are often exceedingly low. The
significance of the correlations between extrinsic and intrinsic noise
also increases when the binding affinity of the signaling molecules
for the detection component increases and/or when the detection
reactions become slower (see Fig.~\ref{fig:noise_pp}). This is
particularly important when weak signals have to be detected, as for
example by the bacterium {\em Escherichia coli}, which can sense
chemical attractants at nanomolar concentration~\cite{Berg04}. To be
able to detect a weak signal, the processing network has to amplify
the signal. This could be achieved by increasing the binding affinity
of the signaling molecules. However, since the association rate cannot
be increased beyond the diffusion-limited association rate constant,
the only way to strongly increase the binding constant is to decrease
the dissociation rate. As Fig.~\ref{fig:noise_pp} shows, this will
increase the importance of the correlations between the extrinsic and
intrinsic noise. Correlations between extrinsic and intrinsic noise
could thus impose strong design constraints for networks that have to
detect small numbers of molecules. Lastly, we emphasize that any
mechanism that increases the gain of a processing network, potentially
also amplifies the effect of the correlations between the extrinsic
and intrinsic noise. An interesting case is provided by the flagellar
rotary motors of {\em E. coli}~\cite{Berg04}. Recent experiments have
revealed that the probability of clockwise rotation depends very
steeply on the concentration of the messenger protein; the effective
Hill coefficient is about 10~\cite{Cluzel00}. This means that the
noise in the rotation direction (the output signal of that system) is
likely to be affected by the binding and unbinding of the messenger
protein to the motor~\cite{Bialek05} and thus to the correlations
between the fluctuations in the messenger protein concentration (the
extrinsic noise) and the intrinsic fluctuations of the motor switching
(intrinsic noise).

Finally, we believe that the predictions of our analysis could be
tested experimentally by performing fluorescence resonance energy
transfer (FRET) or fluorescence correlation spectroscopy
experiments on signal transduction
pathways~\cite{Sato02,Sourjik02}. The MAPK cascade would be an
interesting model system. By putting a FRET donor on MEK, and a FRET
acceptor on both the enzyme of the upstream module, Mos,
and that of the downstream module, MAPK, it should be possible to study the
effect of correlations between extrinsic and intrinsic noise on the
transmission of noise in signal transduction cascades. 

We thank Daan Frenkel, 
\mbox{Bela Mulder}, \mbox{Rosalind Allen} and \mbox{Martin Howard} for a
critical reading of the manuscript. This work is supported by the Amsterdam
Centre for Computational Science (ACCS). The work is part of the research
program of the ``Stichting voor Fundamenteel Onderzoek der Materie (FOM)",
which is financially supported by the ``Nederlandse organisatie voor
Wetenschappelijk Onderzoek (NWO)".

\appendix

\section{General analysis}
\label{app:GenAna}
We consider a general chemical network containing $N$ species
$\{\mathrm{X}_1,\mathrm{X}_2,\dots,\mathrm{X}_N\}$. The state of the
network is defined by the vector $\tebf{X} \equiv
\{{X}_1,{X}_2,\dots,{X}_N\}$, where $X_i$ is the copy number of
component $\rm{X_i}$. The $M$ reactions that constitute the network
will be described by a propensity function ${\mathbf{A}}(\tebf{X})$
and a stoichiometric matrix $\tebf{v}^i$, where $\mathrm{v}^i_j$
denotes the change in the copy number of the species ${\mathrm{X}}_j$
due to reaction $i$.  The reaction network dynamics can be modelled
with a Chemical Master Equation, which describes the evolution of the
probability $P(\bol{X},t)$ of having $X_i$ molecules of type
$\mathrm{X}_i$ \beq \frac{\rmd P(\bol{X},t)}{\rmd
t}=\sum_{i=1}^M\left[
A_i(\bol{X}-\bol{v}^i)P(\bol{X}-\bol{v}^i)-A_i(\bol{X})P(\bol{X})
\right]
\label{cme}
\eeq
Following \cite{Gillespie00}, the dynamics of the system can be approximated
in the limit of large number of molecules by a
Chemical Langevin Equation (CLE) of the form: 
\beq
\frac{\rmd \bol{X}}{\rmd t}=\sum_{k=1}^M \bol{v}^k A_k(\bol{X}) + \bol{\xi}
\label{lang}
\eeq where $\bol{\xi}$ are Gaussian white noise terms of zero average
and variance $\avg{\xi_i \xi_j}=\Xi_{ij}=\sum_{k=1}^M
\bol{v}^k_i\bol{v}^k_j A_k(\bol{X})$.  Setting the noise terms to zero
one obtains the ``deterministic rate equation''. The stable solutions
of the equation \beq 0=\sum_{k=1}^M \bol{v}^k A_k(\bol{X}) \eeq are
usually a
good approximation to the average values $\avg{\bol{X}}$ obtained from
Eq.~\ref{cme} in the limit of large volume
\cite{VanKampen92}.  One can further simplify Eq.~\ref{lang} by
linearizing it around these solutions.  One then obtains a set of
linear stochastic differential equations for
$\bol{x}=\bol{X}-\avg{\bol{X}}$. These represent the Linear Noise
Approximation:
 \beq 
\frac{\rmd x_i}{\rmd t}=\sum_{j=1}^N F_{ij} x_j +
\xi_i \quad \forall 0 < i \leq N 
\eeq
where
\beq
F_{ij}=
\sum_{k=1}^M \bol{v}^k_i \frac{\partial A_k(\avg{\bol{X}})}{\partial
X_j} 
\eeq 
In this approximation, the noise characteristics
(correlation matrix $\bol{\Xi}$) do not depend anymore on the
dynamical variables (here, $\bol{x}$). The fluctuations of the
variables $X_i$ are given by the correlation matrix $\bol{C}$, where
$C_{ij}=\avg{x_i x_j}$. This correlation matrix can be obtained from
the matrix equation~\cite{Gardinerbook} \beq \bol{F C}+\bol{C F}^\dag
=-\bol{\Xi}
\label{FDT}
\eeq The method that we have used to estimate the magnitude of the
fluctuations of the components in the networks discussed below,
consists of two steps: first, we set the noise terms to zero in the
CLE to obtain the steady-state solutions of the resulting
deterministic rate equation; second, we compute the force matrix
$\bol{F}$ and the noise correlations $\bol{\Xi}$ in these points in
order to obtain the correlation matrix $\bol{C}$. In what follows, we
will refer to the value of the matrix element $C_{ii}$ as the noise in
component $\rm{X_i}$. We have verified the accuracy of the Linear
Noise Approximation for the problems discussed here by comparing the
results with those of kinetic Monte Carlo simulations of the chemical
master equation (Eq.~\ref{cme})~\cite{Bortz75,Gillespie77}.\\

\section{General Processing Network and Modularity}
\label{app:GenProc}

Let us consider a linear(ized) reaction network of $N$ components using the
following set of Chemical Langevin Equations: 
\begin{equation}
\dot x_i = \sum_{j=1}^M G_{ij} s_j+\sum_{j=1}^N A_{ij} x_j  + \xi_i \quad\forall\,i = 1
\dots N.  \label{eq:genlin}
\end{equation}
Here, $x_j$ are the components of the processing network and $s_j$
denote the input signals. $A_{ij}$ are coefficients, $G_{ij}$ are
transmission factors, and $\bol{\xi}$ model the noise in the network
components; we model them as Gaussian white noise of zero average and
with $\avg{\xi_i(t)\xi_j(t')}=\Xi_{ij}\delta(t-t')$.  Crucially, in
general, $\bol{s}$ depends upon the \emph{history} of the values of
the network components. Using the linearity of Eq.~\ref{eq:genlin}, the
Fourier transform of the outputs ${\mathrm{X}}_i$ can be found from an
algebraic equation with solution
\begin{equation}
\tilde x_i(\omega)=-\sum_{j=1}^N B_{ij} \left[\sum_{k=1}^M (G_{jk} \tilde s_k(\omega)) + \tilde
\xi_j(\omega)\right],
\end{equation}
where the matrix $\mathbf{B}$ is given by $\mathbf{B}=\left( \mathbf{A} + 
\text{i} \omega \right)^{-1}$. The corresponding power spectrum is:
\begin{widetext}
\begin{eqnarray}
S_{\mathrm{X}_i}(\omega) =\sum_{k,j=1}^N B^\dag_{ki}(\omega)B_{ij}(\omega) \Xi_{jk} + 
\sum_{j,l=1}^N\sum_{k,m=1}^M B_{ij}(\omega)G_{jk} \avg{\tilde s_k(\omega) 
\tilde s^*_m(\omega)}G^\dag_{ml}B^\dag_{li}(\omega) + \nn +
\sum_{j,l=1}^N\sum_{k=1}^M B_{ij}(\omega)
\left[G_{jk}C_{\mathrm{S}_k\xi_l}(\omega)+ G_{lk}C^*_{\mathrm{S}_k\xi_j}(\omega)\right]
B^\dag_{li}(\omega).
\label{eq:genS}
\end{eqnarray}
\end{widetext}
Here, $\tilde C_{\mathrm{S} \xi} (\omega)$ denotes the correlation
between the input signal and the noise $\xi$ in the frequency
domain. The first term of the rhs of Eq.~\ref{eq:genS} is the power
spectrum of the intrinsic noise of ${\mathrm{X}}_i$, while the second
term is the spectrum of the signal $s_j$ modulated by an
\emph{intrinsic} transfer function. If the input signals are
uncorrelated with each other ($\avg{\tilde s_k(\omega) \tilde
s^*_m(\omega)}=\delta_{km}S_{\mathrm{S}_k}(\omega)$) and uncorrelated
with the detection network ($\tilde C_{\mathrm{S} \xi} (\omega) =0$),
Eq.~\ref{eq:genS} reduces to \beq
S_{\mathrm{X}_i}(\omega)=S_{\mathrm{X}_i}^{\mathrm{in}}(\omega)+
\sum_{j=1}^M g_{\mathrm{X}_i}^{\mathrm{S}_j} S_{\mathrm{S}_k}(\omega),
\eeq where $S_{\mathrm{X}_i}^{\mathrm{in}}(\omega)=\sum_{k,j=1}^N
B^\dag_{ki}(\omega)B_{ij}(\omega) \Xi_{jk}$ is the intrinsic noise and
$g_{\mathrm{X}_i}^{\mathrm{S}_k}=\sum_{j,l=1}^N
B_{ij}(\omega)G_{jk}G^\dag_{kl}B^\dag_{li}(\omega)$ are
\emph{intrinsic} transfer functions, both independent of the input
signals. This is the spectral addition rule (Eq.7 of the manuscript)
for a general linear(ized) reaction network that detects multiple
input signals via detection scheme III. Importantly, in this case the
power spectra of the input signals $S_{{\rm S}_i} (\omega) =
\langle\tilde s_i(\omega) \tilde s^*_i(\omega)\rangle$ are unaffected
by their interactions with the processing network; conversely, the
intrinsic contribution to the power spectrum of ${\rm X}_i$,
$S_{\mathrm{X}_i}^{\mathrm{in}}(\omega)$, is a truly intrinsic
quantity that depends upon properties of the processing module only,
and {\em not} upon the fluctuations in the input signals, given by
$S_{\mathrm{S}_k}(\omega)$. However, if the input signals are
correlated with each other and/or detected via scheme I and/or II,
then the power spectra of the input signals and those of the network
do mutually affect each other and the spectral addition rule breaks
down. We emphasize that even when the input signals do not directly
interact with the components that provide the output signals (as in
the section~\ref{sec:CorExIn}), but only
indirectly via chemical reactions of the type of scheme I and II, then
cross-correlations in the noise $\avg{\xi_i \xi_j}$ can be important,
because their effects can propagate from the input signals to the
output signals.  From the perspective of noise transmission, a network
can be decomposed into modules by cutting the network links that
correspond to chemical reactions of the type in scheme III.\\

\begin{widetext}
\section{Zero-Order Ultrasensitive push-pull module }
\label{app:pp}
Let us consider the network:
\vspace*{1 em}
\begin{table}[h]
\begin{tabular}{c|c}
\hline\\[-6pt]
\includegraphics[width=2cm]{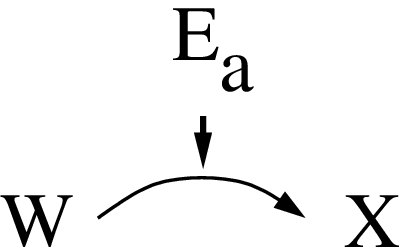}&
$\mathrm{W}+\mathrm{E_a} \equil{a_1}{d_1}\mathrm{E_aW} \give{k_1} \mathrm{E_a+X}$  \\
\hline\\[-6pt]
\includegraphics[width=2cm]{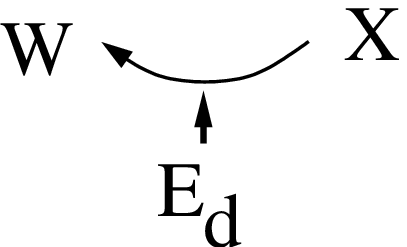}&
$\mathrm{X}+\mathrm{E_d} \equil{a_2}{d_2}\mathrm{E_dX} \give{k_2} \mathrm{E_d+W}$   \\
\hline
\end{tabular}
\label{supptab1}
\end{table}

\noindent The input signal is provided by the enzyme $\mathrm{E_a}$. Its
activation and deactivation dynamics is modeled as a birth-death process:
\beq
\nothing \give{k} \mathrm{E_a} \give{\lambda} \nothing
\eeq
Michaelis-Menten formulae can be derived by assuming fast
equilibration of the complexes $\mathrm{E_aW}$ and $\mathrm{E_dX}$. In
order to self-consistently eliminate these intermediates, we split
the independent variables into slow variables -- those that do not
change in the fast (un)binding reactions -- and fast variables.  We thus
define the following slow variables:
$\mathrm{W_T}=\mathrm{W}+\mathrm{E_aW}$,
$\mathrm{X_T}=\mathrm{X}+\mathrm{E_dX}$,
$\mathrm{E_{a,T}}=\mathrm{E_a}+\mathrm{E_aW}$ and
$\mathrm{E_{d,T}}=\mathrm{E_d}+\mathrm{E_dX}$. The independent
variables therefore are: 
\begin{equation} 
\tebf{X} = \{\tebf{X}_s,\tebf{X}_f\}, \quad
\tebf{X}_s = \{E_{a,T},X_T\}, \quad
\tebf{X}_f=\{E_aW,E_dX\}. 
\end{equation}
We also use the
conservation laws $W+E_aW+X+E_dX=Y = \mbox{constant}$ and
$E_d+E_dX=E_{d,T}$. The values
of the constants $Y$ and $E_{d,T}$ will appear as parameters in the
dynamics of the network.  For the full network, in the new variables,
we have 

\beq \bol{F}=\begin{bmatrix} -\lambda&0&\lambda&0\\
0&0&k_1&-k_2\\ a_1 \avg{W}&-a_1\avg{E_a}&-d_1-k_1-a_1\avg{W+E_a}&0\\
0&a_2 E_d&0&-d_2-k_2-a_2\avg{X-E_d}
\end{bmatrix} \mbox{, and}
\eeq
\beq
\bol{\Xi}=\begin{bmatrix}
k+\lambda \avg{E_a} &0&0&0\\
0&k_1\avg{ E_a}+k_2 \avg{E_dX} &-k_1 \avg{E_aW}&k_2 \avg{E_dX}\\
0&-k_1 \avg{E_aW}&(d_1+k_1)\avg{E_aW}+a_1\avg{E_a}\avg{W}&0\\
0&k_2 \avg{E_dX}&0&(d_2+k_2)\avg{E_dX}+a_2\avg{E_d}\avg{X}
\end{bmatrix}
\eeq

In the limit of large (un)binding rates ($a,d \gg k$), the copy numbers
of the enzyme-substrate
complexes can be expressed as a function of those of the slow
variables

\bea
\label{eq:EaWEdX}
E_aW(X_T, E_{a,T})& =&\frac{ {E_{a,T}} +  {K_{\mathrm{m}1}}+ (Y - X_T) - 
         {\sqrt{{{\left({E_{a,T}} + {K_{\mathrm{m}1}} + Y-X_T \right) }^2} - 
             4\, {E_{a,T}}\left( Y -X_T\right)}}}{2}\nn
E_dX(X_T, E_{a,T})&=&\frac{ {E_{d,T}} +  {K_{\mathrm{m}2}} + X_T - 
         {\sqrt{{{\left( {E_{d,T}} +  {K_{\mathrm{m}2}} + X_T \right) }^2} - 
             4\, {E_{d,T}}\,X_T}}}{2}.
\eea

It is seen that the concentrations of the complexes depend on the rates
$a_i$, $d_i$ and $k_i$ only through the Michaelis-Menten constants
$K_{\mathrm{m}i}=(d_i+k_i)/a_i$. The network dynamics can be reduced to the simple form of a 
two-dimensional Chemical Langevin Equation, depending only on $K_{\mathrm{m}i}$ and $k_i$:
\begin{subequations}
\begin{align}
\frac{\rmd X_{T}}{\rmd t}&= k_1 \, E_aW(X_T, E_{a,T}) - k_2 \, E_dX(X_T,E_{a,T}) +\xi_{\mathrm{X_T}} \\
\frac{\rmd E_{a,T}}{\rmd t} &= k-\lambda \, \left[E_{a,T}- E_aW(X_T,
  E_{a,T})\right]+\xi_{\mathrm{E_{a,T}}} \label{eq:dEaTdt}
\end{align}
\end{subequations}
In order to compute the noise in the network components within the Linear Noise Approximation we use the matrices
\beq
\bol{F}=\begin{bmatrix}
-\lambda+\lambda \frac{\partial E_aW(X_T, E_{a,T})}{\partial E_{a,T}} &\lambda \frac{\partial E_aW(X_T, E_{a,T})}{\partial X_{T}}  \\
k_1 \frac{\partial E_aW(X_T, E_{a,T})}{\partial E_{a,T}}-k_2 \frac{\partial E_dX(X_T, E_{a,T})}{\partial E_{a,T}} & 
k_1 \frac{\partial E_aW(X_T, E_{a,T})}{\partial X_{T}}-k_2 \frac{\partial E_dX(X_T, E_{a,T})}{\partial X_{T}}\\
\end{bmatrix}
\eeq
and 
\beq
\bol{\Xi}=\begin{bmatrix}
k+\lambda \avg{E_a} &0\\
0&k_1\avg{ E_a}+k_2 \avg{E_dX}\\
\end{bmatrix}.
\eeq
The prediction of the noise addition rule for the noise in
$\mathrm{X_T}$ is obtained by setting the feedback $F_{1,2}$ to zero,
such that

\beq
\bol{F_{NA}}=\begin{bmatrix}
-\lambda+\lambda \frac{\partial E_aW(X_T, E_{a,T})}{\partial E_{a,T}} &0 \\
k_1 \frac{\partial E_aW(X_T, E_{a,T})}{\partial E_{a,T}}-k_2 \frac{\partial E_dX(X_T, E_{a,T})}{\partial E_{a,T}} & 
k_1 \frac{\partial E_aW(X_T, E_{a,T})}{\partial X_{T}}-k_2 \frac{\partial E_dX(X_T, E_{a,T})}{\partial X_{T}}\\
\end{bmatrix}.
\eeq

It is perhaps instructive to compare the approach presented
here to that commonly employed for enzymatic
reactions~\cite{Detwiler00,Berg00,Shibata05}, which would write the
reactions for the push-pull network as:
\begin{equation}
\mathrm{W} \equil{k_a}{k_d} \mathrm{X};
    \,\, k_a = \frac{k_1 E_{a,T} \, W}{W +
    K_{\mathrm{m}1}}; \,\, k_d = \frac{k_2 E_{d,T} \, X}{X + K_{\mathrm{m}2}}
\end{equation}
A connection can be made by noting that, when the binding and
unbinding of the enzymes to the substrates is fast, $E_aW = E_{a,T}
W/(K_{{\rm m},1} + W)$ and $E_dX = E_{d,T} X/(K_{{\rm m},2} + X)$; it
can be verified that these expressions are equivalent to those for
$E_aW$ and $E_dX$ in Eq.~\ref{eq:EaWEdX}. However, the principal difference
between our approach and that presented
in~\cite{Detwiler00,Berg00,Shibata05} is that we analyze the dynamics
of $E_{aT}$ and $X_T$ together, thus taking into the fact that the
fluctuations in $X_T$ can act back on those in
$E_{aT}$. If the fraction of enzyme $\rm{E_a}$ that is bound to its
substrate, $\rm{W}$, is small, then the importance of the third term in
Eq.~\ref{eq:dEaTdt} (involving the complex $E_aW$) is small. In this
limit, the results of the full analysis, which takes into account the
correlations between the extrinsic and intrinsic noise, reduces to
those of the spectral addition rule (see Fig.~\ref{fig:noise_pp}).\\

\begin{table*}[t]
\begin{tabular}{|c|}
\hline
$\mathrm{Mos}+\mathrm{E_1} \equil{a_1}{d_1}\mathrm{E_1Mos} \give{k_1} \mathrm{E_1+Mos^*}$ \\
\hline
$\mathrm{Mos^*}+\mathrm{E_2} \equil{a_2}{d_2}\mathrm{E_2Mos^*} \give{k_2} \mathrm{E_2+Mos}$ \\
\hline
$\mathrm{MEK}+\mathrm{Mos^*} \equil{a_3}{d_3}\mathrm{Mos^*MEK} \give{k_3} \mathrm{Mos^*+MEK\!\!\!-\!\!\!P}$\\
\hline
$\mathrm{MEK\!\!\!-\!\!\!P}+\mathrm{MEKP'ase} \equil{a_4}{d_4}\mathrm{MEKP'aseMEK\!\!\!-\!\!\!P} \give{k_4} \mathrm{MEKP'ase+MEK}$ \\
\hline
$\mathrm{MEK\!\!\!-\!\!\!P}+\mathrm{Mos^*} \equil{a_5}{d_5}\mathrm{Mos^*MEK\!\!\!-\!\!\!P} \give{k_5} \mathrm{Mos^*+MEK\!\!\!-\!\!\!PP}$ \\
\hline
$\mathrm{MEK\!\!\!-\!\!\!PP}+\mathrm{MEKP'ase} \equil{a_6}{d_6}\mathrm{MEKP'aseMEK\!\!\!-\!\!\!PP} \give{k_5} \mathrm{MEKP'ase+MEK\!\!\!-\!\!\!P}$ \\
\hline
$\mathrm{MAPK}+\mathrm{MEK\!\!\!-\!\!\!PP} \equil{a_7}{d_7}\mathrm{MEK\!\!\!-\!\!\!PP MAPK} \give{k_7} \mathrm{MEK\!\!\!-\!\!\!PP+MAPPK\!\!\!-\!\!\!P}$ \\
\hline
$\mathrm{MAPK\!\!\!-\!\!\!P}+\mathrm{MAPKP'ase} \equil{a_8}{d_8}\mathrm{MAPKP'aseMAPK\!\!\!-\!\!\!P} \give{k_8} \mathrm{MAPKP'ase+MAPK}$ \\
\hline
$\mathrm{MAPK\!\!\!-\!\!\!P}+\mathrm{MEK\!\!\!-\!\!\!PP} \equil{a_9}{d_9}\mathrm{MEK\!\!\!-\!\!\!PP MAPK\!\!\!-\!\!\!P} \give{k_9} \mathrm{MEK\!\!\!-\!\!\!PP+MAPPK\!\!\!-\!\!\!PP}$\\
\hline
$\mathrm{MAPK\!\!\!-\!\!\!PP}+\mathrm{MAPKP'ase} \equil{a_{10}}{d_{10}}\mathrm{MAPKP'aseMAPK\!\!\!-\!\!\!PP} \give{k_{10}} \mathrm{MAPKP'ase+MAPK\!\!\!-\!\!\!P}$ \\
\hline
\end{tabular}
\caption{Reactions in the MAPK cascade.}
\label{tab:MAPKc}
\end{table*}

\section{MAPK cascade}
\label{app:MAPK}
We model the Mos/MEK/p42 MAPK cascade as a network consisting of 10
enzymatic reactions \cite{Huang96}; these are listed in Table~\ref{tab:MAPKc}.

Since the association and deassociation rates ($a_i$ and $d_i$,
respectively) have not been measured experimentally, we consider the
limit of fast binding and unbinding. We are thus interested in the
dynamics of the slow components: 

\bea 
\mathrm{Mos^*_T} &\equiv &
\mathrm{Mos^*}+\mathrm{E_2Mos^*}+\mathrm{Mos^*MEK}+\mathrm{Mos^*MEK\!\!\!-\!\!\!P}\nn
\mathrm{MEK\!\!\!-\!\!\!P_T}&\equiv &
\mathrm{MEK\!\!\!-\!\!\!P}+\mathrm{Mos^*MEK\!\!\!-\!\!\!P}+\mathrm{MEKP'aseMEK\!\!\!-\!\!\!P}\nn
\mathrm{MEK\!\!\!-\!\!\!PP_T}&\equiv &
\mathrm{MEK\!\!\!-\!\!\!PP}+\mathrm{MEKP'aseMEK\!\!\!-\!\!\!PP}+
\mathrm{MEK\!\!\!-\!\!\!PPMAPK}+\mathrm{MEK\!\!\!-\!\!\!PP MAPK\!\!\!-\!\!\!P}\nn
\mathrm{MAPK\!\!\!-\!\!\!P_T}&\equiv &
\mathrm{MAPK\!\!\!-\!\!\!P}+\mathrm{MEK\!\!\!-\!\!\!PPMAPK\!\!\!-\!\!\!P}+
\mathrm{MAPKP'aseMAPK\!\!\!-\!\!\!P}\nn
\mathrm{MAPK\!\!\!-\!\!\!PP_T}&\equiv &\mathrm{MAPK\!\!\!-\!\!\!PP}+\mathrm{MAPKP'aseMAPK\!\!\!-\!\!\!PP}
\eea
We use the conservation laws:
\bea
\mathrm{E_{1,T}} &=& \mathrm{E_1}+\mathrm{E_1Mos}= \mathrm{constant}\nn
\mathrm{Mos_T} &=& \mathrm{Mos^*_T}+\mathrm{Mos}+\mathrm{E_1Mos}=\mathrm{constant}\nn
\mathrm{E_{2,T}}&=&\mathrm{E_2}+\mathrm{E_2Mos^*}=\mathrm{constant}\nn
\mathrm{MEK_T} &=& \mathrm{MEK\!\!\!-\!\!\!P_T}+\mathrm{MEK\!\!\!-\!\!\!PP_T}+\mathrm{MEK}+\mathrm{Mos^*MEK}=\mathrm{constant}\nn
\mathrm{MEKP'ase_T}&=&\mathrm{MEKP'ase}+\mathrm{MEKP'aseMEK\!\!\!-\!\!\!P}+\mathrm{MEKP'aseMEK\!\!\!-\!\!\!PP}=\mathrm{constant}\nn
\mathrm{MAPK_T}&=&\mathrm{MAPK}+\mathrm{MAPK\!\!\!-\!\!\!P_T}+\mathrm{MAPK\!\!\!-\!\!\!PP_T}+\mathrm{MEK\!\!\!-\!\!\!PP
MAPK}=\mathrm{constant}\\
\mathrm{MAPKP'ase_T} &=&\mathrm{MAPKP'ase}+\mathrm{MAPKP'aseMAPK\!\!\!-\!\!\!P}+\mathrm{MAPKP'aseMAPK\!\!\!-\!\!\!PP}=\mathrm{constant}\nonumber
\label{total}
\eea
For the slow variables, the Chemical Langevin Equations read:
\bea
\frac{\rmd}{\rmd t} Mos^*_T&=&k_1  E_1Mos - k_2 E_2Mos^*+\xi_1 \nn
\frac{\rmd}{\rmd t} MEK\!\!\!-\!\!\!P_T&=&k_3 Mos^*MEK -k_4 MEKP'aseMEK\!\!\!-\!\!\!P- \nn
&&k_5 Mos^*MEK\!\!\!-\!\!\!P+k_6 MEKP'aseMEK\!\!\!-\!\!\!PP+\xi_2\nn
\frac{\rmd}{\rmd t} MEK\!\!\!-\!\!\!PP_T&=&k_5 Mos^*MEK\!\!\!-\!\!\!P-k_6 MEKP'aseMEK\!\!\!-\!\!\!PP+\xi_3\nn
\frac{\rmd}{\rmd t} MAPK\!\!\!-\!\!\!P_T&=&k_7 {MEK\!\!\!-\!\!\!PP MAPK}-k_8 {MAPKP'aseMAPK\!\!\!-\!\!\!P}-\nn
&&k_9 MEK\!\!\!-\!\!\!PP MAPK\!\!\!-\!\!\!P + k_{10} MAPKP'aseMAPK\!\!\!-\!\!\!PP+\xi_4 \nn
\frac{\rmd}{\rmd t} {MAPK\!\!\!-\!\!\!PP_T}&=&k_9 MEK\!\!\!-\!\!\!PP MAPK\!\!\!-\!\!\!P-k_{10} MAPKP'aseMAPK\!\!\!-\!\!\!PP+\xi_5
\label{CLEMAPK}
\eea
where the noise terms $\bol{\xi}$, as discussed above, are modeled as
Gaussian white noise of zero mean and variance:
\bea
\label{eq:MAPKNOISE_DIAG}
\avg{\xi_1^2}&=&k_1  E_1Mos + k_2 E_2Mos^* \nn
\avg{\xi_2^2}&=&k_3 Mos^*MEK + k_4 MEKP'aseMEK\!\!\!-\!\!\!P+ \nn
&&k_5 Mos^*MEK\!\!\!-\!\!\!P+k_6 MEKP'aseMEK\!\!\!-\!\!\!PP\nn
\avg{\xi_3^2}&=&k_5 Mos^*MEK\!\!\!-\!\!\!P+k_6 MEKP'aseMEK\!\!\!-\!\!\!PP\nn
\avg{\xi_4^2}&=&k_7 {MEK\!\!\!-\!\!\!PP MAPK}+k_8 {MAPKP'aseMAPK\!\!\!-\!\!\!P}+\nn
&&k_9 MEK\!\!\!-\!\!\!PP MAPK\!\!\!-\!\!\!P + k_{10} MAPKP'aseMAPK\!\!\!-\!\!\!PP \nn
\avg{\xi_5^2}&=&k_9 MEK\!\!\!-\!\!\!PP MAPK\!\!\!-\!\!\!P+k_{10} MAPKP'aseMAPK\!\!\!-\!\!\!PP
\eea
All the noise terms are uncorrelated, except:
\bea
\label{eq:MAPKNOISE_OFFDIAG}
\avg{\xi_2\xi_3}=- \left(k_5 Mos^*MEK\!\!\!-\!\!\!P+k_6 MEKP'aseMEK\!\!\!-\!\!\!PP\right) \\
\avg{\xi_4\xi_5}= -\left(k_9 MEK\!\!\!-\!\!\!PP MAPK\!\!\!-\!\!\!P + k_{10} MAPKP'aseMAP\right)
\eea

The noise correlations $\avg{\xi_i \xi_j}$ define the elements of the
matrix $\bol{\Xi}$.

The dependence of the intermediate complexes $\mathrm{E_1Mos}$,
$\mathrm{E_2Mos^*}$, $\mathrm{Mos^*MEK}$,
$\mathrm{Mos^*MEK\!\!\!-\!\!\!P}$,
$\mathrm{MEKP'aseMEK\!\!\!-\!\!\!P}$,
$\mathrm{MEKP'aseMEK\!\!\!-\!\!\!PP}$, $\mathrm{MEK\!\!\!-\!\!\!PP
  MAPK}$, $\mathrm{MEK\!\!\!-\!\!\!PP MAPK\!\!\!-\!\!\!P}$,
$\mathrm{MAPKP'aseMAPK\!\!\!-\!\!\!P}$, and
$\mathrm{MAPKP'aseMAPK\!\!\!-\!\!\!PP}$ on the slow variables around
the steady state,
 has been obtained numerically.  The concentrations of the complexes depend on the reaction rates only 
through  the Michaelis-Menten constants $K_{\mathrm{m}i}=(d_i+k_i)/a_i$.

We can now construct, as explained in section~\ref{app:GenAna}, the force
 matrix $\bol{F}$ of the Linear Noise approximation and use
 $\bol{F}$ and $\bol{\Xi}$ as defined in Eqs.~\ref{eq:MAPKNOISE_DIAG}
and~\ref{eq:MAPKNOISE_OFFDIAG}  to numerically solve
 Eq. \ref{FDT}.

The MAPK network consists of three layers. Here we address the
question to what extent these can be described as three independent
modules. To this end, we divide the full network into subnetworks, in
three different manners:

\bea \mathrm{1.}  & \{\mathrm{Mos^*_T}\} &
\{\mathrm{MEK\!\!\!-\!\!\!P_T},\mathrm{MEK\!\!\!-\!\!\!PP_T}\} \quad
\{\mathrm{MAPK\!\!\!-\!\!\!P_T},\mathrm{MAPK\!\!\!-\!\!\!PP_T}\} \nn
\mathrm{2.}  &
\{\mathrm{Mos^*_T},\mathrm{MEK\!\!\!-\!\!\!P_T},\mathrm{MEK\!\!\!-\!\!\!PP_T}\}&
\{\mathrm{MAPK\!\!\!-\!\!\!P_T},\mathrm{MAPK\!\!\!-\!\!\!PP_T}\} \nn
\mathrm{3.}  & \{\mathrm{Mos^*_T}\} &
\{\mathrm{MEK\!\!\!-\!\!\!P_T},\mathrm{MEK\!\!\!-\!\!\!PP_T},\mathrm{MAPK\!\!\!-\!\!\!P_T},\mathrm{MAPK\!\!\!-\!\!\!PP_T}\}
\eea The network numbered 1 consists of three uncoupled layers; that
numbered 2 consists of one module in which the Mos and the MEK layer
are coupled, while the MAPK layer is assumed to form a second,
independent module; network 3 consists of one module in which layers 2
and 3 are concatenated and one, uncoupled, module formed by layer
1.

To study the effect of the correlations between the extrinsic and
intrinsic noise, it is instructive to define submatrices,
$\bf{F_{i i}}$, $\bf{G_{ij}}$ ($i>j$), and $\bf{K_{i j}}$ ($i<j$) of the
matrix $\bf{F}$ of the full network. These matrices are defined as follows:
\beq
\begin{bmatrix}
\bol{F_{11}} & \bol{K_{12}} &\bol{K_{13}}\\
\bol{G_{21}} & \bol{F_{22}} &\bol{K_{23}}\\
\bol{G_{31}} & \bol{G_{32}} &\bol{F_{33}}\\
\end{bmatrix}\equiv \bol{F}.
\eeq

The three decompositions of the full network then correspond to the
following force matrices:
\begin{align}
\mbox{1}:&\mbox{ Spectral Addition; All Uncoupled   }&\hspace*{.5cm} \bol{F_1}\equiv\begin{bmatrix}
\bol{F_{11}} & \bol{0} &\bol{0}\\
\bol{G_{21}} & \bol{F_{22}} &\bol{0}\\
\bol{0} & \bol{G_{32}} &\bol{F_{33}}\;\\
\end{bmatrix}\nn
\mbox{2}:&\mbox{ Coupled Mos and MEK; Uncoupled MAPK}&\hspace*{.5cm}\bol{F_2}\equiv\begin{bmatrix}
\bol{F_{11}} & \bol{K_{12}} &\bol{0}\\
\bol{G_{21}} & \bol{F_{22}} &\bol{0}\\
\bol{0} & \bol{G_{32}} &\bol{F_{33}}\;\\
\end{bmatrix}\nn
\mbox{3}:&\mbox{ Uncoupled Mos; Coupled MEK and MAPK}&\hspace*{.5cm}\bol{F_3}\equiv\begin{bmatrix}
\bol{F_{11}} & \bol{0} &\bol{0}\\
\bol{G_{21}} & \bol{F_{22}} &\bol{K_{23}}\\
\bol{0} & \bol{G_{32}} &\bol{F_{33}}\\
\end{bmatrix}\nn
\end{align}
Here, $\bol{0}$ is the null matrix.

\begin{table}[t]
\begin{tabular}{|c|c|c|c|c|c|c|}
\hline
$\mathrm{Mos_T}$&$\mathrm{MEK_T}$&$\mathrm{MAPK_T}$&
$\quad\mathrm{E_{1,T}}\quad$& 
$\quad\mathrm{E_{2,T}}\quad$& 
$\mathrm{MEKP'ase_{T}}$& 
$\mathrm{MAPKP'ase_T}$\\
\hline
3 &1200&300&0.1&0.6&0.6&300 \\
\hline
\end{tabular}
\caption{Total concentrations of the components (nM)}
\label{totalc}
\end{table}

The noise matrix $\bol{\Xi}$ of the full network is already
partitioned into the different modules. For all partitionings, the
noise matrix is thus given by Eqs.~\ref{eq:MAPKNOISE_DIAG}
and~\ref{eq:MAPKNOISE_OFFDIAG}. 

The relevant model parameters are the total concentrations on the rhs
of Eq.~\ref{total} and the Michaelis-Menten constants. The concentrations
were taken from the experiments discussed in~\cite{Ferrell96,
Huang96, Angeli04, Mansour96} and are shown in
Table~\ref{totalc}. The Michaelis-Menten constant of the MAPK
activation by MEK has been measured
experimentally~\cite{Mansour96}. Following~\cite{Huang96,Angeli04}, we take all
Michaelis-Menten constants to be equal to $K_{\rm mi} = K_{\rm M} = 300
\,$ nM. The $k_i$ values were taken to be equal to each other; their
absolute value is not important for the calculations here, because it
only sets the time scale in the problem.

\end{widetext}


\begin{thebibliography}{45}
\expandafter\ifx\csname natexlab\endcsname\relax\def\natexlab#1{#1}\fi
\expandafter\ifx\csname bibnamefont\endcsname\relax
  \def\bibnamefont#1{#1}\fi
\expandafter\ifx\csname bibfnamefont\endcsname\relax
  \def\bibfnamefont#1{#1}\fi
\expandafter\ifx\csname citenamefont\endcsname\relax
  \def\citenamefont#1{#1}\fi
\expandafter\ifx\csname url\endcsname\relax
  \def\url#1{\texttt{#1}}\fi
\expandafter\ifx\csname urlprefix\endcsname\relax\def\urlprefix{URL }\fi
\providecommand{\bibinfo}[2]{#2}
\providecommand{\eprint}[2][]{\url{#2}}

\bibitem[{\citenamefont{McAdams and Arkin}(1997)}]{McAdams97}
\bibinfo{author}{\bibfnamefont{H.~H.} \bibnamefont{McAdams}} \bibnamefont{and}
  \bibinfo{author}{\bibfnamefont{A.}~\bibnamefont{Arkin}},
  \bibinfo{journal}{Proc. Natl. Acad. Sci. USA} \textbf{\bibinfo{volume}{94}},
  \bibinfo{pages}{814 } (\bibinfo{year}{1997}).

\bibitem[{\citenamefont{Elowitz and Leibler}(2000)}]{Elowitz00}
\bibinfo{author}{\bibfnamefont{M.~B.} \bibnamefont{Elowitz}} \bibnamefont{and}
  \bibinfo{author}{\bibfnamefont{S.}~\bibnamefont{Leibler}},
  \bibinfo{journal}{Nature} \textbf{\bibinfo{volume}{403}}, \bibinfo{pages}{335
  } (\bibinfo{year}{2000}).

\bibitem[{\citenamefont{Ozbudak et~al.}(2002)\citenamefont{Ozbudak, Thattai,
  Kurtser, Grossman, and van Oudenaarden}}]{Ozbudak02}
\bibinfo{author}{\bibfnamefont{E.~M.} \bibnamefont{Ozbudak}},
  \bibinfo{author}{\bibfnamefont{M.}~\bibnamefont{Thattai}},
  \bibinfo{author}{\bibfnamefont{I.}~\bibnamefont{Kurtser}},
  \bibinfo{author}{\bibfnamefont{A.~D.} \bibnamefont{Grossman}},
  \bibnamefont{and} \bibinfo{author}{\bibfnamefont{A.}~\bibnamefont{van
  Oudenaarden}}, \bibinfo{journal}{Nat. Gen.} \textbf{\bibinfo{volume}{31}},
  \bibinfo{pages}{69 } (\bibinfo{year}{2002}).

\bibitem[{\citenamefont{Elowitz et~al.}(2002)\citenamefont{Elowitz, Levine,
  Siggia, and Swain}}]{Elowitz02}
\bibinfo{author}{\bibfnamefont{M.~B.} \bibnamefont{Elowitz}},
  \bibinfo{author}{\bibfnamefont{A.~J.} \bibnamefont{Levine}},
  \bibinfo{author}{\bibfnamefont{E.~D.} \bibnamefont{Siggia}},
  \bibnamefont{and} \bibinfo{author}{\bibfnamefont{P.~S.} \bibnamefont{Swain}},
  \bibinfo{journal}{Science} \textbf{\bibinfo{volume}{297}},
  \bibinfo{pages}{1183 } (\bibinfo{year}{2002}).

\bibitem[{\citenamefont{Ozbudak et~al.}(2004)\citenamefont{Ozbudak, Thattai,
  Lim, Shraiman, and van Oudenaarden}}]{Ozbudak04}
\bibinfo{author}{\bibfnamefont{E.~M.} \bibnamefont{Ozbudak}},
  \bibinfo{author}{\bibfnamefont{M.}~\bibnamefont{Thattai}},
  \bibinfo{author}{\bibfnamefont{H.~N.} \bibnamefont{Lim}},
  \bibinfo{author}{\bibfnamefont{B.~I.} \bibnamefont{Shraiman}},
  \bibnamefont{and} \bibinfo{author}{\bibfnamefont{A.}~\bibnamefont{van
  Oudenaarden}}, \bibinfo{journal}{Nature} \textbf{\bibinfo{volume}{427}},
  \bibinfo{pages}{737} (\bibinfo{year}{2004}).

\bibitem[{\citenamefont{Pedraza and van Oudenaarden}(2005)}]{Pedraza05}
\bibinfo{author}{\bibfnamefont{J.~M.} \bibnamefont{Pedraza}} \bibnamefont{and}
  \bibinfo{author}{\bibfnamefont{A.}~\bibnamefont{van Oudenaarden}},
  \bibinfo{journal}{Science} \textbf{\bibinfo{volume}{307}},
  \bibinfo{pages}{1965 } (\bibinfo{year}{2005}).

\bibitem[{\citenamefont{Rosenfeld et~al.}(2005)\citenamefont{Rosenfeld, Young,
  Alon, Swain, and Elowitz}}]{Rosenfeld05}
\bibinfo{author}{\bibfnamefont{N.}~\bibnamefont{Rosenfeld}},
  \bibinfo{author}{\bibfnamefont{J.~W.} \bibnamefont{Young}},
  \bibinfo{author}{\bibfnamefont{U.}~\bibnamefont{Alon}},
  \bibinfo{author}{\bibfnamefont{P.~S.} \bibnamefont{Swain}}, \bibnamefont{and}
  \bibinfo{author}{\bibfnamefont{M.~B.} \bibnamefont{Elowitz}},
  \bibinfo{journal}{Science} \textbf{\bibinfo{volume}{307}},
  \bibinfo{pages}{1962} (\bibinfo{year}{2005}).

\bibitem[{\citenamefont{Blake et~al.}(2003)\citenamefont{Blake, Kaern, Cantor,
  and Collins}}]{Blake03}
\bibinfo{author}{\bibfnamefont{W.~J.} \bibnamefont{Blake}},
  \bibinfo{author}{\bibfnamefont{M.}~\bibnamefont{Kaern}},
  \bibinfo{author}{\bibfnamefont{C.~R.} \bibnamefont{Cantor}},
  \bibnamefont{and} \bibinfo{author}{\bibfnamefont{J.~J.}
  \bibnamefont{Collins}}, \bibinfo{journal}{Nature}
  \textbf{\bibinfo{volume}{422}}, \bibinfo{pages}{633} (\bibinfo{year}{2003}).

\bibitem[{\citenamefont{Raser and O'Shea}(2004)}]{Raser04}
\bibinfo{author}{\bibfnamefont{J.~M.} \bibnamefont{Raser}} \bibnamefont{and}
  \bibinfo{author}{\bibfnamefont{E.~K.} \bibnamefont{O'Shea}},
  \bibinfo{journal}{Science} \textbf{\bibinfo{volume}{304}},
  \bibinfo{pages}{1811} (\bibinfo{year}{2004}).

\bibitem[{\citenamefont{Acar et~al.}(2005)\citenamefont{Acar, Becksei, and van
  Oudenaarden}}]{Acar05}
\bibinfo{author}{\bibfnamefont{M.}~\bibnamefont{Acar}},
  \bibinfo{author}{\bibfnamefont{A.}~\bibnamefont{Becksei}}, \bibnamefont{and}
  \bibinfo{author}{\bibfnamefont{A.}~\bibnamefont{van Oudenaarden}},
  \bibinfo{journal}{Nature} \textbf{\bibinfo{volume}{435}},
  \bibinfo{pages}{228} (\bibinfo{year}{2005}).

\bibitem[{\citenamefont{Becksei et~al.}(2005)\citenamefont{Becksei, Kaufmann,
  and van Oudenaarden}}]{Becksei05}
\bibinfo{author}{\bibfnamefont{A.}~\bibnamefont{Becksei}},
  \bibinfo{author}{\bibfnamefont{B.~B.} \bibnamefont{Kaufmann}},
  \bibnamefont{and} \bibinfo{author}{\bibfnamefont{A.}~\bibnamefont{van
  Oudenaarden}}, \bibinfo{journal}{Nat. Genet.} \textbf{\bibinfo{volume}{37}},
  \bibinfo{pages}{937} (\bibinfo{year}{2005}).

\bibitem[{\citenamefont{Colman-Lerner et~al.}(2005)\citenamefont{Colman-Lerner,
  Gordon, Serra, Chin, Resnekov, Endy, Pesce, and Brent}}]{ColmanLerner05}
\bibinfo{author}{\bibfnamefont{A.}~\bibnamefont{Colman-Lerner}},
  \bibinfo{author}{\bibfnamefont{A.}~\bibnamefont{Gordon}},
  \bibinfo{author}{\bibfnamefont{E.}~\bibnamefont{Serra}},
  \bibinfo{author}{\bibfnamefont{T.}~\bibnamefont{Chin}},
  \bibinfo{author}{\bibfnamefont{O.}~\bibnamefont{Resnekov}},
  \bibinfo{author}{\bibfnamefont{D.}~\bibnamefont{Endy}},
  \bibinfo{author}{\bibfnamefont{G.}~\bibnamefont{Pesce}}, \bibnamefont{and}
  \bibinfo{author}{\bibfnamefont{R.}~\bibnamefont{Brent}},
  \bibinfo{journal}{Nature} \textbf{\bibinfo{volume}{437}},
  \bibinfo{pages}{699} (\bibinfo{year}{2005}).

\bibitem[{\citenamefont{Korobkova et~al.}(2004)\citenamefont{Korobkova, Emonet,
  Vilar, Shimizu, and Cluzel}}]{Korobkova04}
\bibinfo{author}{\bibfnamefont{E.}~\bibnamefont{Korobkova}},
  \bibinfo{author}{\bibfnamefont{T.}~\bibnamefont{Emonet}},
  \bibinfo{author}{\bibfnamefont{J.~M.~G.} \bibnamefont{Vilar}},
  \bibinfo{author}{\bibfnamefont{T.~S.} \bibnamefont{Shimizu}},
  \bibnamefont{and} \bibinfo{author}{\bibfnamefont{P.}~\bibnamefont{Cluzel}},
  \bibinfo{journal}{Nature} \textbf{\bibinfo{volume}{428}},
  \bibinfo{pages}{574} (\bibinfo{year}{2004}).

\bibitem[{\citenamefont{Rao et~al.}(2002)\citenamefont{Rao, Wolf, and
  Arkin}}]{Rao02}
\bibinfo{author}{\bibfnamefont{C.~V.} \bibnamefont{Rao}},
  \bibinfo{author}{\bibfnamefont{D.~M.} \bibnamefont{Wolf}}, \bibnamefont{and}
  \bibinfo{author}{\bibfnamefont{A.~P.} \bibnamefont{Arkin}},
  \bibinfo{journal}{Nature} \textbf{\bibinfo{volume}{420}}, \bibinfo{pages}{231
  } (\bibinfo{year}{2002}).

\bibitem[{\citenamefont{Kaern et~al.}(2005)\citenamefont{Kaern, Elston, Blake,
  and Collins}}]{Kaern05}
\bibinfo{author}{\bibfnamefont{M.}~\bibnamefont{Kaern}},
  \bibinfo{author}{\bibfnamefont{T.~C.} \bibnamefont{Elston}},
  \bibinfo{author}{\bibfnamefont{W.~J.} \bibnamefont{Blake}}, \bibnamefont{and}
  \bibinfo{author}{\bibfnamefont{J.~J.} \bibnamefont{Collins}},
  \bibinfo{journal}{Nat. Rev. Genet.} \textbf{\bibinfo{volume}{6}},
  \bibinfo{pages}{451 } (\bibinfo{year}{2005}).

\bibitem[{\citenamefont{Paulsson et~al.}(2000)\citenamefont{Paulsson, Berg, and
  Ehrenberg}}]{Paulsson00_1}
\bibinfo{author}{\bibfnamefont{J.}~\bibnamefont{Paulsson}},
  \bibinfo{author}{\bibfnamefont{O.~G.} \bibnamefont{Berg}}, \bibnamefont{and}
  \bibinfo{author}{\bibfnamefont{M.}~\bibnamefont{Ehrenberg}},
  \bibinfo{journal}{Proc. Natl. Acad. Sci. USA} \textbf{\bibinfo{volume}{97}},
  \bibinfo{pages}{7148 } (\bibinfo{year}{2000}).

\bibitem[{\citenamefont{Vilar et~al.}(2002)\citenamefont{Vilar, Kueh, Barkai,
  and Leibler}}]{Vilar02}
\bibinfo{author}{\bibfnamefont{J.~M.~G.} \bibnamefont{Vilar}},
  \bibinfo{author}{\bibfnamefont{H.~Y.} \bibnamefont{Kueh}},
  \bibinfo{author}{\bibfnamefont{N.}~\bibnamefont{Barkai}}, \bibnamefont{and}
  \bibinfo{author}{\bibfnamefont{S.}~\bibnamefont{Leibler}},
  \bibinfo{journal}{Proc. Natl. Acad. Sci. USA} \textbf{\bibinfo{volume}{99}},
  \bibinfo{pages}{5988 } (\bibinfo{year}{2002}).

\bibitem[{\citenamefont{Thattai and van Oudenaarden}(2004)}]{Thattai04}
\bibinfo{author}{\bibfnamefont{M.}~\bibnamefont{Thattai}} \bibnamefont{and}
  \bibinfo{author}{\bibfnamefont{A.}~\bibnamefont{van Oudenaarden}},
  \bibinfo{journal}{Genetics} \textbf{\bibinfo{volume}{167}},
  \bibinfo{pages}{523} (\bibinfo{year}{2004}).

\bibitem[{\citenamefont{Kussell and Leibler}(2005)}]{Kussell05}
\bibinfo{author}{\bibfnamefont{E.}~\bibnamefont{Kussell}} \bibnamefont{and}
  \bibinfo{author}{\bibfnamefont{S.}~\bibnamefont{Leibler}},
  \bibinfo{journal}{Science} \textbf{\bibinfo{volume}{309}},
  \bibinfo{pages}{2075} (\bibinfo{year}{2005}).

\bibitem[{\citenamefont{Hartwell et~al.}(1999)\citenamefont{Hartwell, Hopfield,
  Leibler, and Murray}}]{Hartwell99}
\bibinfo{author}{\bibfnamefont{L.~H.} \bibnamefont{Hartwell}},
  \bibinfo{author}{\bibfnamefont{J.~J.} \bibnamefont{Hopfield}},
  \bibinfo{author}{\bibfnamefont{S.}~\bibnamefont{Leibler}}, \bibnamefont{and}
  \bibinfo{author}{\bibfnamefont{A.~W.} \bibnamefont{Murray}},
  \bibinfo{journal}{Nature} \textbf{\bibinfo{volume}{402}},
  \bibinfo{pages}{C47} (\bibinfo{year}{1999}).

\bibitem[{\citenamefont{Kashtan and Alon}(2005)}]{Kashtan05}
\bibinfo{author}{\bibfnamefont{N.}~\bibnamefont{Kashtan}} \bibnamefont{and}
  \bibinfo{author}{\bibfnamefont{U.}~\bibnamefont{Alon}},
  \bibinfo{journal}{Proc. Natl. Acad. Sci. USA} \textbf{\bibinfo{volume}{102}},
  \bibinfo{pages}{13773} (\bibinfo{year}{2005}).

\bibitem[{\citenamefont{Angeli et~al.}(2004)\citenamefont{Angeli, Ferrell, Jr,
  and Sontag}}]{Angeli04}
\bibinfo{author}{\bibfnamefont{D.}~\bibnamefont{Angeli}},
  \bibinfo{author}{\bibfnamefont{J.~E.} \bibnamefont{Ferrell}},
  \bibinfo{author}{\bibnamefont{Jr}}, \bibnamefont{and}
  \bibinfo{author}{\bibfnamefont{E.~D.} \bibnamefont{Sontag}},
  \bibinfo{journal}{PNAS} \textbf{\bibinfo{volume}{101}}, \bibinfo{pages}{1822}
  (\bibinfo{year}{2004}).

\bibitem[{\citenamefont{Detwiler et~al.}(2000)\citenamefont{Detwiler,
  Ramanathan, Sengupta, and Shraimann}}]{Detwiler00}
\bibinfo{author}{\bibfnamefont{P.~B.} \bibnamefont{Detwiler}},
  \bibinfo{author}{\bibfnamefont{S.}~\bibnamefont{Ramanathan}},
  \bibinfo{author}{\bibfnamefont{A.}~\bibnamefont{Sengupta}}, \bibnamefont{and}
  \bibinfo{author}{\bibfnamefont{B.~I.} \bibnamefont{Shraimann}},
  \bibinfo{journal}{Biophys. J.} \textbf{\bibinfo{volume}{79}},
  \bibinfo{pages}{2801 } (\bibinfo{year}{2000}).

\bibitem[{\citenamefont{Thattai and van Oudenaarden}(2002)}]{Thattai02}
\bibinfo{author}{\bibfnamefont{M.}~\bibnamefont{Thattai}} \bibnamefont{and}
  \bibinfo{author}{\bibfnamefont{A.}~\bibnamefont{van Oudenaarden}},
  \bibinfo{journal}{Biophys. J.} \textbf{\bibinfo{volume}{82}},
  \bibinfo{pages}{2943 } (\bibinfo{year}{2002}).

\bibitem[{\citenamefont{Paulsson}(2004)}]{Paulsson04}
\bibinfo{author}{\bibfnamefont{J.}~\bibnamefont{Paulsson}},
  \bibinfo{journal}{Nature} \textbf{\bibinfo{volume}{427}}, \bibinfo{pages}{415
  } (\bibinfo{year}{2004}).

\bibitem[{\citenamefont{Shibata and Fujimoto}(2005)}]{Shibata05}
\bibinfo{author}{\bibfnamefont{T.}~\bibnamefont{Shibata}} \bibnamefont{and}
  \bibinfo{author}{\bibfnamefont{K.}~\bibnamefont{Fujimoto}},
  \bibinfo{journal}{Proc. Natl. Acad. Sci. USA} \textbf{\bibinfo{volume}{102}},
  \bibinfo{pages}{331 } (\bibinfo{year}{2005}).

\bibitem[{\citenamefont{Goldbeter and {Koshland, Jr.}}(1981)}]{Goldbeter81}
\bibinfo{author}{\bibfnamefont{A.}~\bibnamefont{Goldbeter}} \bibnamefont{and}
  \bibinfo{author}{\bibfnamefont{D.~E.} \bibnamefont{{Koshland, Jr.}}},
  \bibinfo{journal}{Proc. Natl. Acad. Sci. USA} \textbf{\bibinfo{volume}{78}},
  \bibinfo{pages}{6840} (\bibinfo{year}{1981}).

\bibitem[{\citenamefont{{Ferrell, Jr.}}(1996)}]{Ferrell96}
\bibinfo{author}{\bibfnamefont{J.~E.} \bibnamefont{{Ferrell, Jr.}}},
  \bibinfo{journal}{Trends Biochem. Sci.} \textbf{\bibinfo{volume}{21}},
  \bibinfo{pages}{460} (\bibinfo{year}{1996}).

\bibitem[{\citenamefont{Berg et~al.}(2000)\citenamefont{Berg, Paulsson, and
  Ehrenberg}}]{Berg00}
\bibinfo{author}{\bibfnamefont{O.~G.} \bibnamefont{Berg}},
  \bibinfo{author}{\bibfnamefont{J.}~\bibnamefont{Paulsson}}, \bibnamefont{and}
  \bibinfo{author}{\bibfnamefont{M.}~\bibnamefont{Ehrenberg}},
  \bibinfo{journal}{Biophys. J.} \textbf{\bibinfo{volume}{79}},
  \bibinfo{pages}{1228} (\bibinfo{year}{2000}).

\bibitem[{\citenamefont{Samoilov et~al.}(2005)\citenamefont{Samoilov,
  Plyasunov, and Arkin}}]{Samoilov05}
\bibinfo{author}{\bibfnamefont{M.}~\bibnamefont{Samoilov}},
  \bibinfo{author}{\bibfnamefont{S.}~\bibnamefont{Plyasunov}},
  \bibnamefont{and} \bibinfo{author}{\bibfnamefont{A.~P.} \bibnamefont{Arkin}},
  \bibinfo{journal}{Proc. Natl. Acad. Sci. USA} \textbf{\bibinfo{volume}{102}},
  \bibinfo{pages}{2310} (\bibinfo{year}{2005}).

\bibitem[{\citenamefont{van Kampen}(1992)}]{VanKampen92}
\bibinfo{author}{\bibfnamefont{N.~G.} \bibnamefont{van Kampen}},
  \emph{\bibinfo{title}{Stochastic Processes in Physics and Chemistry}}
  (\bibinfo{publisher}{North-Holland}, \bibinfo{address}{Amsterdam},
  \bibinfo{year}{1992}).

\bibitem[{\citenamefont{Simpson et~al.}(2004)\citenamefont{Simpson, Cox, and
  Sayler}}]{Simpson04}
\bibinfo{author}{\bibfnamefont{M.~L.} \bibnamefont{Simpson}},
  \bibinfo{author}{\bibfnamefont{C.~D.} \bibnamefont{Cox}}, \bibnamefont{and}
  \bibinfo{author}{\bibfnamefont{G.~S.} \bibnamefont{Sayler}},
  \bibinfo{journal}{J. Theor. Biol.} \textbf{\bibinfo{volume}{229}},
  \bibinfo{pages}{383 } (\bibinfo{year}{2004}).

\bibitem[{\citenamefont{Felder et~al.}(1992)\citenamefont{Felder, LaVin,
  Ullrich, and Schlessinger}}]{Felder92}
\bibinfo{author}{\bibfnamefont{S.}~\bibnamefont{Felder}},
  \bibinfo{author}{\bibfnamefont{J.}~\bibnamefont{LaVin}},
  \bibinfo{author}{\bibfnamefont{A.}~\bibnamefont{Ullrich}}, \bibnamefont{and}
  \bibinfo{author}{\bibfnamefont{J.}~\bibnamefont{Schlessinger}},
  \bibinfo{journal}{J. Cell Biol.} \textbf{\bibinfo{volume}{117}},
  \bibinfo{pages}{202} (\bibinfo{year}{1992}).

\bibitem[{\citenamefont{Gillespie}(2000)}]{Gillespie00}
\bibinfo{author}{\bibfnamefont{D.~T.} \bibnamefont{Gillespie}},
  \bibinfo{journal}{J. Chem. Phys.} \textbf{\bibinfo{volume}{113}},
  \bibinfo{pages}{297 } (\bibinfo{year}{2000}).

\bibitem[{\citenamefont{Bortz et~al.}(1975)\citenamefont{Bortz, Kalos, and
  Lebowitz}}]{Bortz75}
\bibinfo{author}{\bibfnamefont{A.~B.} \bibnamefont{Bortz}},
  \bibinfo{author}{\bibfnamefont{M.~H.} \bibnamefont{Kalos}}, \bibnamefont{and}
  \bibinfo{author}{\bibfnamefont{J.~L.} \bibnamefont{Lebowitz}},
  \bibinfo{journal}{J. Comp. Phys.} \textbf{\bibinfo{volume}{17}},
  \bibinfo{pages}{10} (\bibinfo{year}{1975}).

\bibitem[{\citenamefont{Gillespie}(1977)}]{Gillespie77}
\bibinfo{author}{\bibfnamefont{D.~T.} \bibnamefont{Gillespie}},
  \bibinfo{journal}{J. Phys. Chem.} \textbf{\bibinfo{volume}{81}},
  \bibinfo{pages}{2340 } (\bibinfo{year}{1977}).

\bibitem[{\citenamefont{Bialek and Setayeshgar}(2005)}]{Bialek05}
\bibinfo{author}{\bibfnamefont{W.}~\bibnamefont{Bialek}} \bibnamefont{and}
  \bibinfo{author}{\bibfnamefont{S.}~\bibnamefont{Setayeshgar}},
  \bibinfo{journal}{Proc. Natl. Acad. Sci. USA} \textbf{\bibinfo{volume}{102}},
  \bibinfo{pages}{10040} (\bibinfo{year}{2005}).

\bibitem[{\citenamefont{Huang and {Ferrell, Jr.}}(1996)}]{Huang96}
\bibinfo{author}{\bibfnamefont{C.-Y.~F.} \bibnamefont{Huang}} \bibnamefont{and}
  \bibinfo{author}{\bibfnamefont{J.~E.} \bibnamefont{{Ferrell, Jr.}}},
  \bibinfo{journal}{Proc. Natl. Acad. Sci. USA} \textbf{\bibinfo{volume}{93}},
  \bibinfo{pages}{10078} (\bibinfo{year}{1996}).

\bibitem[{\citenamefont{{Ferrell, Jr.} and Machleder}(1996)}]{Ferrell98}
\bibinfo{author}{\bibfnamefont{J.~E.} \bibnamefont{{Ferrell, Jr.}}}
  \bibnamefont{and} \bibinfo{author}{\bibfnamefont{E.~M.}
  \bibnamefont{Machleder}}, \bibinfo{journal}{Proc. Natl. Acad. Sci. USA}
  \textbf{\bibinfo{volume}{93}}, \bibinfo{pages}{10078} (\bibinfo{year}{1996}).

\bibitem[{\citenamefont{Berg}(2004)}]{Berg04}
\bibinfo{author}{\bibfnamefont{H.~C.} \bibnamefont{Berg}},
  \emph{\bibinfo{title}{E. coli in motion}} (\bibinfo{publisher}{Springer},
  \bibinfo{address}{New York}, \bibinfo{year}{2004}).

\bibitem[{\citenamefont{Cluzel et~al.}(2000)\citenamefont{Cluzel, Surette, and
  Leibler}}]{Cluzel00}
\bibinfo{author}{\bibfnamefont{P.}~\bibnamefont{Cluzel}},
  \bibinfo{author}{\bibfnamefont{M.}~\bibnamefont{Surette}}, \bibnamefont{and}
  \bibinfo{author}{\bibfnamefont{S.}~\bibnamefont{Leibler}},
  \bibinfo{journal}{Science} \textbf{\bibinfo{volume}{287}},
  \bibinfo{pages}{1652} (\bibinfo{year}{2000}).

\bibitem[{\citenamefont{Sato et~al.}(2002)\citenamefont{Sato, Ozawa, Inukai,
  Asano, and Umezawa}}]{Sato02}
\bibinfo{author}{\bibfnamefont{M.}~\bibnamefont{Sato}},
  \bibinfo{author}{\bibfnamefont{T.}~\bibnamefont{Ozawa}},
  \bibinfo{author}{\bibfnamefont{K.}~\bibnamefont{Inukai}},
  \bibinfo{author}{\bibfnamefont{T.}~\bibnamefont{Asano}}, \bibnamefont{and}
  \bibinfo{author}{\bibfnamefont{Y.}~\bibnamefont{Umezawa}},
  \bibinfo{journal}{Nat. Biotech.} \textbf{\bibinfo{volume}{20}},
  \bibinfo{pages}{287} (\bibinfo{year}{2002}).

\bibitem[{\citenamefont{Sourjik and Berg}(2002)}]{Sourjik02}
\bibinfo{author}{\bibfnamefont{V.}~\bibnamefont{Sourjik}} \bibnamefont{and}
  \bibinfo{author}{\bibfnamefont{H.~C.} \bibnamefont{Berg}},
  \bibinfo{journal}{Proc. Natl. Acad. Sci. USA} \textbf{\bibinfo{volume}{99}},
  \bibinfo{pages}{123 } (\bibinfo{year}{2002}).

\bibitem[{\citenamefont{Gardiner}(2004)}]{Gardinerbook}
\bibinfo{author}{\bibfnamefont{C.~W.} \bibnamefont{Gardiner}},
  \emph{\bibinfo{title}{Handbook of Stochastic Methods, 3rd edition}}
  (\bibinfo{publisher}{Springer-Verlag}, \bibinfo{address}{Berlin},
  \bibinfo{year}{2004}).

\bibitem[{\citenamefont{Mansour et~al.}(1996)\citenamefont{Mansour, Candia,
  Matsuura, Manning, and Ahn}}]{Mansour96}
\bibinfo{author}{\bibfnamefont{S.}~\bibnamefont{Mansour}},
  \bibinfo{author}{\bibfnamefont{J.}~\bibnamefont{Candia}},
  \bibinfo{author}{\bibfnamefont{J.}~\bibnamefont{Matsuura}},
  \bibinfo{author}{\bibfnamefont{M.}~\bibnamefont{Manning}}, \bibnamefont{and}
  \bibinfo{author}{\bibfnamefont{N.}~\bibnamefont{Ahn}},
  \bibinfo{journal}{Biochemistry} \textbf{\bibinfo{volume}{35}},
  \bibinfo{pages}{15529} (\bibinfo{year}{1996}).


\end{thebibliography}
\end{document}